\documentclass[12pt,onecolumn,draftclsnofoot]{IEEEtran}

\usepackage{setspace}
\usepackage{graphicx}
\usepackage{cite}
\usepackage{algorithmic}
\usepackage{algorithm}
\usepackage{amssymb}
\usepackage{subfigure}
\usepackage[cmex10]{amsmath}
\usepackage{url}
\begin{document}
%
% paper title
% can use linebreaks \\ within to get better formatting as desired
\title{Adaptive Cut Generation Algorithm for Improved Linear Programming Decoding of Binary Linear Codes}

\author{Xiaojie~Zhang%\IEEEmembership{,~Student Member,~IEEE}%
        ~and~Paul~H.~Siegel\IEEEmembership{, Fellow,~IEEE}% <-this % stops a space
\thanks{This research was supported in part by the Center for Magnetic Recording Research at the University of California, San Diego and by the National Science Foundation under Grant CCF-0829865.}%
\thanks{The material in this paper was presented in part at the IEEE International Symposium on Information Theory, Saint Petersburg, Russia, July 31 -- August 5, 2011.}%
\thanks{Xiaojie Zhang and Paul H. Siegel are with the Department of Electrical and Computer Engineering and the Center for Magnetic Recording Research, University of California, San Diego, La Jolla, CA 92093 (emails: \{ericzhang, psiegel\}@ucsd.edu)}}% <-this % stops a space
\markboth{Submitted to IEEE Trans. Inform. Theory}{}
% make the title area
\maketitle

\begin{abstract}
%\boldmath
Linear programming (LP) decoding approximates maximum-likelihood (ML) decoding of a linear block code by relaxing the equivalent ML integer programming (IP) problem into a more easily solved LP problem. The LP problem is defined by a set of box constraints together with a set of linear inequalities called ``parity inequalities'' that are derived from the constraints represented by the rows of a parity-check matrix of the code and can be added iteratively and adaptively. In this paper, we first derive a new necessary condition and a new sufficient condition for a violated parity inequality constraint, or ``cut,'' at a point in the unit hypercube. Then, we propose a new and effective algorithm to generate parity inequalities derived from certain additional redundant parity check (RPC) constraints that can eliminate pseudocodewords produced by the LP decoder, often significantly improving the decoder error-rate performance. The cut-generating algorithm is based upon a specific transformation of an initial parity-check matrix of the linear block code. We also design two variations of the proposed decoder to make it more efficient when it is combined with the new cut-generating algorithm. Simulation results for several low-density parity-check (LDPC) codes demonstrate that the proposed decoding algorithms significantly narrow the performance gap between LP decoding and ML decoding.
\end{abstract}

% Note that keywords are not normally used for peerreview papers.

\IEEEpeerreviewmaketitle

\section{Introduction}
\label{sec:intro}
\IEEEPARstart{L}{ow}-density parity-check (LDPC) codes were first introduced by Gallager in the 1960s \cite{Gallager}, together with a class of iterative decoding algorithms. Later, in the 1990s, the rediscovery of LDPC codes by MacKay and Neal \cite{MacKayNeal1995}, \cite{MacKayNeal1997} launched a period of intensive research on these codes and their decoding algorithms. Significant attention was paid to iterative message-passing (MP) decoders, particularly belief propagation (BP) \cite{BP} as embodied by the sum-product algorithm (SPA) \cite{SPA}.

Despite the unparalleled success of iterative decoding in practice, it is quite difficult to analyze the performance of such iterative MP decoders due to the heuristic nature of their message update rules and their local nature. An alternative approach, linear programming (LP) decoding, was introduced by Feldman et al. \cite{Feldman_LP} as an approximation to maximum-likelihood (ML) decoding.

Many theoretical and empirical observations suggest similarities between the performance of LP and MP decoding methods. For example, graph-cover decoding can be used as a theoretical tool to show the connection between LP decoding and iterative MP decoding~\cite{Vontobel_GC}.

However, there are some key differences that distinguish LP decoding from iterative MP decoding. One of these differences is that the LP decoder has the \emph{ML certificate property}, i.e., it is detectable if the decoding algorithm fails to find an ML codeword. When it fails to find an ML codeword, the LP decoder finds a non-integer solution, commonly called a \emph{pseudocodeword}. Another difference is that while adding redundant parity checks satisfied by all the codewords can only improve LP decoding, adding redundant parity checks may have a negative effect on MP decoding, especially in the waterfall region, due to the creation of short cycles in the Tanner graph. This property of LP decoding allows improvements by tightening the LP relaxation, i.e., reducing the feasible space of the LP problem by adding more linear constraints from redundant parity checks.

In the original formulation of LP decoding proposed by Feldman \emph{et al.}, the number of constraints in the LP problem is linear in the block-length but exponential in the maximum check node degree, and the authors also argued that the number of useful constraints could be reduced to polynomial in code length. The computational complexity of the original LP formulation therefore can be prohibitively high, motivating the design of computationally simplified decoding algorithms that can achieve the same error-rate performance with a smaller number of constraints.
For example, efficient polynomial-time algorithms can be used for solving the original LP formulation \cite{FeldmanThesis}. An alternative LP formulation whose size is linear in the check node degree and code length can also be obtained by changing the graphical representation of the code \cite{YangFeldman,Dendro}; namely, all check nodes of high degree are replaced by dendro-subgraphs (trees) with an appropriate number of auxiliary degree-3 check nodes and degree-2 variable nodes. Several other low-complexity LP decoders were also introduced in \cite{lowcpxLP}, suggesting that LP solvers with complexity similar to the min-sum algorithm and the sum-product algorithm are feasible.

Another approach is to add linear constraints in an adaptive and selective way during the LP formulation~\cite{Taghavi_ALP}. Such an adaptive linear programming (ALP) decoding approach also allows the adaptive incorporation of linear constraints generated by redundant parity checks (RPC) into the LP problem, making it possible to reduce the feasible space and improve the system performance. A linear inequality derived from an RPC that eliminates a pseudocodeword solution is referred to as a ``cut.''

An algorithm proposed in \cite{Taghavi_ALP} uses a random walk on a subset of the code factor graph to find these RPC cuts. However, the random nature of this algorithm limits its efficiency. In fact, experiments show that the average number of random trials required to find an RPC cut grows exponentially with the length of the code.

Recently, the authors of \cite{SepAlg} proposed a separation algorithm that derives Gomory cuts from the IP formulation of the decoding problem and finds cuts from RPCs which are generated by applying Gaussian elimination to the original parity-check matrix.
%, but without any column permutation.
In~\cite{cutting-plane}, a cutting-plane method was proposed to improve the fractional distance of a given binary parity-check matrix -- the minimum weight of nonzero vertices of the fundamental polytope -- by adding redundant rows obtained by converting the parity-check matrix into row echelon form after a certain column permutation. However, we have observed that the RPCs obtained by the approach in~\cite{cutting-plane} are not able to produce enough cuts to improve the error-rate performance relative to the separation algorithm when they are used in conjunction with either ALP decoding or the separation algorithm. A detailed survey on mathematical programming approaches for decoding binary linear codes can be found in \cite{LPsurvey}.

In this paper, we greatly improve the error-correcting performance of LP decoding by designing algorithms that can efficiently generate cut-inducing RPCs and find possible cuts from such RPCs. First, we derive a new necessary condition and a new sufficient condition for a parity-check to provide a cut at a given pseudocodeword. These conditions naturally suggest an efficient algorithm that can be used to find, for a given pseudocodeword solution to an LP problem,  the unique cut (if it exists) among the parity inequalities associated with a parity check. This algorithm was previously introduced by Taghavi \emph{et al.}~\cite[Algorithm~2]{SparseLP} and, independently and in a slightly different form, by Wadayama~\cite[Fig.~6]{Wadayama}.

The conditions also serve as the motivation for a new, more efficient adaptive cut-inducing RPC generation algorithm that identifies useful RPCs by performing specific elementary row operations on the original parity-check matrix of the binary linear code. By adding the corresponding linear constraints into the LP problem, we can significantly improve the error-rate performance of the LP decoder, even approaching the ML decoder performance in the high-SNR region for some codes. Finally, we modify the ALP decoder to make it more efficient when being combined with the new cut-generating algorithm. Simulation results demonstrate that the proposed decoding algorithms significantly improve the error-rate performance of the original LP decoder.

The remainder of the paper is organized as follows. In Section~\ref{sec:LPD}, we review the original formulation of LP decoding and several adaptive LP decoding algorithms. Section~\ref{sec:ECSA} presents the new necessary condition and new sufficient condition for a parity-check to induce a cut, as well as their connection to the efficient cut-search algorithm. In Section~\ref{sec:ACGA}, we describe our proposed algorithm for finding RPC-based cuts. Section~\ref{sec:NR} presents our simulation results, and Section~\ref{sec:concl} concludes the paper.

\section{Linear Programming Decoding and Adaptive Variants}
\label{sec:LPD}

\subsection{Linear Programming (LP) Relaxation of Maximum Likelihood (ML) Decoding}
\label{subsec:LPR}

Consider a binary linear block code $\mathcal{C}$ of length $n$ and a corresponding $m\times n$ parity-check matrix $\mathbf{H}$. A codeword $\mathbf{y}\in\mathcal{C}$ is transmitted across a memoryless binary-input  output-symmetric channel, resulting in a received vector $\mathbf{r}$. Assuming that the transmitted codewords are equiprobable, the ML decoder finds the solution to the following optimization problem (see, e.g., \cite{Taghavi_ALP})
\begin{equation}
\label{MLopt}
\begin{array}{l}
 {\text{minimize}\quad}{{\boldsymbol{\gamma}}^T}{\mathbf{u}} \\
 {\text{subject to}\quad}{\mathbf{u}} \in {\mathcal C} \\
 \end{array}
\end{equation}
where $u_i\in\{0,1\}$, and $\boldsymbol{\gamma}$ is the vector of log-likelihood ratios (LLR) defined as
\begin{equation}
\label{LLR}
{\gamma_i} = \log \left( {\frac{\Pr \left( {\left. {R_i = r_i} \right|{u_i} = 0} \right)}{\Pr \left( {\left. {R_i = r_i} \right|{u_i} = 1} \right)}} \right).
\end{equation}

Since the ML decoding problem \eqref{MLopt} is an integer programming problem, it is desirable to replace its integrality constraints with a set of linear constraints, transforming the IP problem into a more readily solved LP problem. The desired feasible space of the corresponding LP problem should be the \emph{codeword polytope}, i.e., the convex hull of all the codewords in $\mathcal{C}$. With this, unless the cost vector of the LP decoding problem is orthogonal to a face of the constraint polytope, the optimal solution is one integral vertex of its codeword polytope, in which case it is the same as the output of the ML decoder. When the LP solution is not unique, there is at least one integral vertex corresponding to an ML codeword.  However, the number of linear constraints typically needed to represent the codeword polytope increases exponentially with the code length, which makes such a relaxation impractical.

As an approximation to ML decoding, Feldman \emph{et al.} \cite{Feldman_LP,FeldmanThesis} relaxed the codeword polytope to a polytope now known as \emph{fundamental polytope} \cite{Vontobel_GC}, denoted as $\mathcal{P}(\mathbf{H})$, which depends on the parity-check matrix $\mathbf H$.

\newtheorem{def1}{Definition}
\begin{def1}[Fundamental polytope \cite{Vontobel_GC}]
\label{LP}
Let us define
\begin{equation}
\label{localcodeword}
\mathcal{C}_j\triangleq\left\{ {{\mathbf{x}} \in \mathbb{F}^n_2|\langle\mathbf{x},\mathbf{h}_j\rangle=0~(\text{in }\mathbb{F}_2)} \right\}
\end{equation}
where ${\mathbf h}_j$ is the $j$th row of the parity-check matrix $\mathbf H$ and $1\leq j\leq m$. Thus, $\mathcal{C}_j$ is the set of all binary vectors satisfying the $j$th parity-check constraint. We denote by $\text{conv}(\mathcal{C}_j)$ the convex hull of $\mathcal{C}_j$ in $\mathbb{R}^n$, which consists of all possible real convex combinations of the points in $\mathcal{C}_j$, now regarded as points in $\mathbb{R}^n$. The fundamental polytope $\mathcal{P}(\mathbf{H})$ of the parity-check matrix $\mathbf H$ is defined to be the set
\begin{equation}
\label{fp}
{\mathcal{P}}\left( {\mathbf{H}} \right) = \bigcap\limits_{j = 1}^m {{\text{conv}}\left( {{\mathcal{C}}_j} \right)}.
\end{equation}
\end{def1}

Therefore, LP decoding can be written as the following optimization problem:

\begin{equation}
\label{LP_fp}
\begin{array}{l}
 {\text{minimize}\quad}{\boldsymbol{\gamma}^T}{\mathbf{u}} \\
 {\text{subject to}\quad}{\mathbf{u}} \in \mathcal{P}(\mathbf{H}). \\
 \end{array}
\end{equation}

The solution of the above LP problem corresponds to a vertex of the fundamental polytope that minimizes the cost function. Since the fundamental polytope has both integral and nonintegral vertices, with the integral vertices corresponding exactly to the codewords of $\mathcal C$  \cite{Feldman_LP,Vontobel_GC}, if the LP solver outputs an integral solution, it must be a valid codeword and is guaranteed to be an ML solution, which is called the \emph{ML certificate property}. The nonintegral solutions are called pseudocodewords. Since the fundamental polytope is a function of the parity-check matrix $\mathbf H$ used to represent the code $\mathcal C$, different parity-check matrices for $\mathcal C$ may have different fundamental polytopes. Therefore, a given code has many possible LP-based relaxations, and some may be better than others when used for LP decoding.

The fundamental polytope can also be described by a set of linear inequalities, obtained as follows \cite{Feldman_LP}.
First of all, for a point $\mathbf{u}$ within the fundamental polytope, it should satisfy the box constraints such that $0\leq u_i\leq 1$, for $i=1,\dots,n$.
Then, let $\mathcal{N}(j)\subseteq\{1,2,\ldots,n\}$ be the set of neighboring variable nodes of the check node $j$ in the Tanner graph, that is, $\mathcal{N}(j)=\{i:H_{j,i}=1\}$ where $H_{j,i}$ is the element in the $j$th row and $i$th column of the parity-check matrix, $\mathbf{H}$.
For each row  $j=1,\dots,m$ of the parity-check matrix, corresponding to a check node in the associated Tanner graph, the linear inequalities used to form the fundamental polytope $\mathcal{P}(\mathbf{H})$ are given by
\begin{equation}
\label{PI1}
\sum\limits_{i \in \mathcal{V}} \left(1-u_i\right)  + \sum\limits_{i \in \mathcal{N}\left( j \right)\backslash \mathcal{V}} {{u_i}}  \geq 1
\text{,\quad for  all \;}  \mathcal{V}\subseteq \mathcal{N}(j){\text{, with }}\left| \mathcal{V} \right|{\text{ odd}}
\end{equation}
where for a set $\mathcal{X}$, $|\mathcal{X}|$ denotes its cardinality. It is easy to see that \eqref{PI1} is equivalent to
\begin{equation}
\label{PI2}
\sum\limits_{i \in \mathcal{V}} {{u_i}}  - \sum\limits_{i \in \mathcal{N}\left( j \right)\backslash \mathcal{V}} {{u_i}}  \leq \left| \mathcal{V} \right| - 1
\text{,\quad for all \;}  \mathcal{V}\subseteq \mathcal{N}(j){\text{, with }}\left| \mathcal{V} \right|{\text{ odd.}}
\end{equation}
Note that, for each check node $j$, the corresponding inequalities in \eqref{PI1} or \eqref{PI2} and the linear box constraints exactly describe the convex hull of the set $\mathcal{C}_j$.

The linear constraints in \eqref{PI1} (and therefore also \eqref{PI2}) are referred to as \emph{parity inequalities}, which are also known as \emph{forbidden set inequalities} \cite{Feldman_LP}. It can be easily verified that these linear constraints are equivalent to the original parity-check constraints when each $u_i$ takes on binary values only.

\newtheorem{prop1}{Proposition}
\begin{prop1}[Theorem 4 in \cite{Feldman_LP}]
\label{prop1}
The parity inequalities of the form  \eqref{PI1} derived from all rows of the parity-check matrix $\mathbf{H}$ and the box constraints completely describe the fundamental polytope $\mathcal{P}(\mathbf{H})$.
\end{prop1}

With this, LP decoding can also be formulated as follows

\begin{equation}
\label{origLP}
\begin{array}{l}
 {\text{minimize}\quad}{{\boldsymbol{\gamma}}^T}{\mathbf{u}} \\
 {\text{subject to}\quad}0\leq u_i\leq 1, {\; \text{for all } \; } i;\\
\text{\quad\qquad\quad~~}\sum\limits_{i \in\mathcal{V}} \left(1-u_i\right)  + \sum\limits_{i \in \mathcal{N}\left( j \right)\backslash \mathcal{V}} {{u_i}}  \geq 1\\
\text{\qquad\qquad\quad for all \;}j, \mathcal{V}\subseteq \mathcal{N}(j){\text{, with }}\left| \mathcal{V} \right|{\text{ odd.}}
\end{array}
\end{equation}

In the following parts of this paper, we refer to the above formulation of LP decoding problem based on the fundamental polytope of the original parity-check matrix as the \emph{original} LP decoding.

\subsection{Adaptive Linear Programming (ALP) Decoding}
\label{subsec:alp}

In the original formulation of LP decoding presented in \cite{Feldman_LP}, every check node $j$ generates $2^{|\mathcal{N}(j)|-1}$ parity inequalities that are used as linear constraints in the LP problem described in \eqref{origLP}.
The total number of constraints and the complexity of the original LP decoding problem grows exponentially with the maximum check node degree. So, even for binary linear codes with moderate check degrees, the number of constraints in the original LP decoding could be prohibitively large. In the literature, several approaches to reducing the complexity of the original LP formulation  have been described \cite{FeldmanThesis,YangFeldman,Dendro,lowcpxLP,Taghavi_ALP}. We will use adaptive linear programming (ALP) decoding \cite{Taghavi_ALP} as the foundation of the improved LP decoding algorithms presented in later sections. The ALP decoder exploits the structure of the LP decoding problem, reflected in the statement of the following lemma.

\newtheorem{lemma1}{Lemma}
\begin{lemma1}[Theorem 1 in \cite{Taghavi_ALP}]
\label{lemma1}
If at any given point $\mathbf{u}\in[0,1]^n$, one of the parity inequalities introduced by a check node $j$ is violated, the rest of the parity inequalities from this check node are satisfied with strict inequality.
\end{lemma1}

\newtheorem{def2}[def1]{Definition}
\begin{def2}
\label{def_cut}
Given a parity-check node $j$, a set $\mathcal{V} \subseteq \mathcal{N}(j)$ of odd cardinality, and a vector $\mathbf{u}\in[0,1]^n$ such that the corresponding parity inequality of the form \eqref{PI1} or \eqref{PI2} does not hold, we say that the constraint is \emph{violated} or, more succinctly, a \emph{cut} at $\mathbf{u}$.
\footnote{In the terminology of \cite{LPsurvey}, if (\ref{PI2}) does not hold for a pseudocodeword $\mathbf{u}$, then the vector $(\mathbf{r},t) \in \mathbb{R}^n \times \mathbb{R}$, where $r_i=1$ for all $i\in \mathcal{V},$ $r_i=-1$ for
all  $i\in \mathcal{N}(j)\backslash\mathcal{V},$ $r_i=0$ otherwise, and $t=|\mathcal{V}|-1$, is a \emph{valid cut}, separating $\mathbf{u}$ from the codeword polytope.}
\end{def2}

In \cite{Taghavi_ALP}, an efficient algorithm for finding cuts at a vector $\mathbf{u}\in[0,1]^n$ was presented. It relies on the observation that violation of a parity inequality \eqref{PI2} at $\mathbf{u}$ implies that
\begin{equation}
\label{cut1}
|\mathcal{V}|-1<\sum\limits_{i \in \mathcal{V}} {{u_i}}\leq |\mathcal{V}|
\end{equation}
and
\begin{equation}
\label{cut2}
0\leq \sum\limits_{i \in \mathcal{N}\left( j \right)\backslash \mathcal{V}} {{u_i}} < u_v, {\; \text{for all \;}} v\in \mathcal{V}.
\end{equation}
where $\mathcal{V}$ is an odd-sized subset of $\mathcal{N}(j)$.

Given a parity check $j$, the algorithm first puts its neighboring variables in $\mathbf{u}$ into non-increasing order, i.e., $u_{j_1}\geq\dots\geq u_{j_n}$, for $u_{j_i}\in\mathcal{N}(j)$. It then successively considers subsets of odd cardinality having the form $\mathcal{V}=\{u_{j_1},\dots,u_{j_{2k+1}}\}\subseteq \mathcal{N}(j)$, increasing the size of  $\mathcal{V}$ by two each step,  until a cut (if one exists) is found. This algorithm can find a cut among the constraints corresponding to a check node $j$ by examining at most $|\mathcal{N}(j)|/2$ inequalities, rather than exhaustively checking all $2^{|\mathcal{N}(j)|-1}$ inequalities in the original formulation of LP decoding.

The ALP decoding algorithm starts by solving the LP problem with the same objective function as \eqref{MLopt}, but with only the following constraints
\begin{equation}
\label{init_cons}
\left\{ \begin{gathered}
  0\leq u_i \quad \text{if}\quad \gamma_i\geq0\hfill\\
  u_i \leq 1\quad \text{if}\quad \gamma_i<0.\hfill \\
\end{gathered}  \right.
\end{equation}
The solution of this initial LP problem can be obtained simply by making a hard decision on the components of a received vector. The ALP decoding algorithm starts with this point, searches every check node for cuts, adds all the cuts found during the search as constraints into the LP problem, and solves it again. This procedure is repeated until an optimal integer solution is generated or no more cuts can be found (see \cite{Taghavi_ALP} for more details). Adaptive LP decoding has exactly the same error-correcting performance as the original LP decoding.

\section{Cut Conditions}
\label{sec:ECSA}

In this section, we derive a necessary condition and a sufficient condition for a parity inequality to be a cut at $\mathbf{u} \in [0,1]^n$. We also show their connection to the
efficient cut-search algorithm proposed by Taghavi \emph{et al.}~\cite[Algorithm~2]{SparseLP} and Wadayama~\cite[Fig.~6]{Wadayama}. This algorithm is more efficient than the search technique from \cite{Taghavi_ALP} that was mentioned in Section~\ref{sec:LPD}.

Consider the original parity inequalities in \eqref{PI1} given by Feldman \emph{et al.} in \cite{Feldman_LP}. If a parity inequality derived from check node $j$ induces a cut at $\mathbf u$, the cut can be written as

\begin{equation}
\label{CUT}
\sum\limits_{i \in \mathcal{V}} \left(1-u_i\right)  + \sum\limits_{i \in \mathcal{N}(j)\backslash \mathcal{V}} {{u_i}}  < 1
,\quad \text{ for some } \mathcal{V}\subseteq \mathcal{N}(j){\text{ with }} |\mathcal{V}| {\text{ odd.}}
\end{equation}

From \eqref{CUT} and Lemma \ref{lemma1}, we can derive the following necessary condition for a parity-check constraint to induce a cut.

\newtheorem{thm1}{Theorem}
\begin{thm1}
\label{thm1}
Given a nonintegral vector $\mathbf{u}$ and a parity check $j$, let $\mathcal{S}=\{i\in\mathcal{N}(j)|0<u_i<1\}$ be the set of nonintegral neighbors of $j$ in the Tanner graph, and let $\mathcal{T}=\{i\in\mathcal{N}(j)|u_i>\frac{1}{2}\}$. A necessary condition for parity check $j$ to induce a cut at $\mathbf{u}$ is
\begin{equation}
\label{CUT_NC}
\sum\limits_{i \in \mathcal{T}} \left(1-u_i\right)  + \sum\limits_{i \in \mathcal{N}(j)\backslash \mathcal{T}} {{u_i}}  < 1.
\end{equation}
This is equivalent to
\begin{equation}
\label{CUT_NCe}
\sum\limits_{i \in \mathcal{S}} \left|\frac{1}{2}-u_i\right| > \frac{1}{2}\cdot|\mathcal{S}|-1
\end{equation}
where, for $x\in\mathbb{R}$, $|x|$ denotes the absolute value.

\end{thm1}
\begin{IEEEproof}
For a given vector $\mathbf{u}$ and a subset $\mathcal{X}\subseteq\mathcal{N}(j)$, define the function
\begin{equation}\notag
\label{gX}
g\left(\mathcal{X}\right)=\sum\limits_{i \in \mathcal{X}} \left(1-u_i\right)  + \sum\limits_{i \in \mathcal{N}\left( j \right)\backslash \mathcal{X}} u_i.
\end{equation}
If parity-check $j$ incudes a cut at $\mathbf{u}$, there must be a set
$\mathcal{V}\subseteq\mathcal{N}(j)$ of odd cardinality such that (\ref{CUT}) holds. This means that $g\left(\mathcal{V}_{\mathrm{cut}}\right)<1$. Now, it is easy to see that the set $\mathcal{T}$ minimizes the function $g\left(\mathcal{X}\right)$, from which it follows that $g\left(\mathcal{T}\right)\leq g\left(\mathcal{V}_{\mathrm{cut}}\right)<1$. Therefore, inequality \eqref{CUT_NC} must hold in order for parity check $j$ to induce a cut.

For $\frac{1}{2}\leq u_i\leq 1$, we have
\begin{equation}\label{uil}\notag
\frac{1}{2}-\left|\frac{1}{2}-u_i\right| = \frac{1}{2}-\left(u_i-\frac{1}{2}\right) = 1-u_i,
\end{equation}
and for $0\leq u_i\leq \frac{1}{2}$, we have
\begin{equation}\label{uis}\notag
\frac{1}{2}-\left|\frac{1}{2}-u_i\right| = \frac{1}{2}-\left(\frac{1}{2}-u_i\right) = u_i.
\end{equation}
Hence, \eqref{CUT_NC} can be rewritten as
\begin{equation}\notag
\sum\limits_{i \in \mathcal{S}} \left(\frac{1}{2}-\left|\frac{1}{2}-u_i\right|\right)<1
\end{equation}
or equivalently,
\begin{equation}\notag
\frac{1}{2}\cdot|\mathcal{S}| - \sum\limits_{i \in \mathcal{S}} \left|\frac{1}{2}-u_i\right|<1\nonumber
\end{equation}
which implies inequality \eqref{CUT_NCe}.
\end{IEEEproof}

\newtheorem{thm1rm}{Remark}
\begin{thm1rm}
\label{thm1rm}
Theorem~\ref{thm1} shows that to see whether a parity-check node could provide a cut at a pseudocodeword $\mathbf{u}$ we only need to examine its fractional neighbors.
\end{thm1rm}

Reasoning similar to that used in the proof of Theorem~\ref{thm1} yields a sufficient condition for a parity-check node to induce a cut at $\mathbf{u}$.

\newtheorem{thm2}[thm1]{Theorem}
\begin{thm2}
\label{thm2}
Given a nonintegral vector $\mathbf{u}$ and a parity check $j$, let $\mathcal{S}=\{i\in\mathcal{N}(j)|0<u_i<1\}$ and $\mathcal{T}=\{i\in\mathcal{N}(j)|u_i>\frac{1}{2}\}$. If the inequality
\begin{equation}
\label{CUT_SC}
\sum\limits_{i \in \mathcal{T}} \left(1-u_i\right)  + \sum\limits_{i \in\mathcal{N}(j) \backslash \mathcal{T}} {{u_i}} + 2\cdot\mathop {\min}\limits_{i\in\mathcal{S}} \left|\frac{1}{2}-u_i\right|  < 1
\end{equation}
holds, there must be a violated parity inequality derived from parity check $j$. This sufficient condition can be written as
\begin{equation}
\label{CUT_SCe}
\sum\limits_{i \in \mathcal{S}} \left|\frac{1}{2}-u_i\right| - 2\cdot\mathop {\min}\limits_{i\in\mathcal{S}} \left|\frac{1}{2}-u_i\right| > \frac{1}{2}\cdot|\mathcal{S}|-1.
\end{equation}
\end{thm2}
\begin{IEEEproof}
Lemma \ref{lemma1} implies that, if parity check $j$ gives a cut at $\mathbf{u}$, then there is at most one odd-sized set $\mathcal{V}\subseteq \mathcal{N}(j)$ that satisfies \eqref{CUT}.
From the proof of Theorem \ref{thm1}, we have $g\left(\mathcal{T}\right)\leq g\left(\mathcal{X}\right)$  $\text{for all \;} \mathcal{X}\subseteq\mathcal{N}(j)$.
If $\left|\mathcal{T}\right|$ is even, we need to find one element $i^*\in\mathcal{N}(j)$ such that inserting it into or removing it from $\mathcal{T}$ would result in the minimum increment to the value of $g\left(\mathcal{T}\right)$. Obviously, $i^* = \arg\mathop {\min}\limits_{i\in\mathcal{N}(j)} \left|\frac{1}{2}-u_i\right|$, and the increment is $2\cdot\left|\frac{1}{2}-u_{i^*}\right|$. If more than one $i$ minimizes the expression $\left|\frac{1}{2}-u_i\right| $, we choose one arbitrarily as $i^*$.  Hence, setting
\begin{equation}
\mathcal{V}=\left\{ \begin{gathered}
  \mathcal{T}\backslash\{i^*\}$, \quad~ if $i^*\in\mathcal{T}\hfill\\
  \mathcal{T}\cup\{i^*\}$, \quad if $i^*\notin\mathcal{T}\nonumber\hfill\\
\end{gathered}  \right.
\end{equation}
we have $g\left(\mathcal{V}\right)= g\left(\mathcal{T}\right) + 2\cdot\left|\frac{1}{2}-u_{i^*}\right| \geq g\left(\mathcal{T}\right)$.
If inequality \eqref{CUT_SC} holds, then $g\left(\mathcal{T}\right)\leq g\left(\mathcal{V}\right)<1$. Since either $|\mathcal{T}|$ or $|\mathcal{V}|$ is odd, \eqref{CUT_SC} is a sufficient condition for parity-check constraint $j$ to induce a cut at $\mathbf{u}$. Arguing as in the latter part of the proof of Theorem~\ref{thm1}, it can be shown that \eqref{CUT_SC} is equivalent to \eqref{CUT_SCe}.
\end{IEEEproof}

Theorem \ref{thm1} and Theorem \ref{thm2} provide a necessary condition and a sufficient condition, respectively, for a parity-check node to produce a cut at any given vector $\mathbf{u}$. It is worth pointing out that \eqref{CUT_NC} becomes a necessary and sufficient condition for a parity check to produce a cut when $|\mathcal{T}|$ is odd, and \eqref{CUT_SC} becomes a necessary and sufficient condition when $|\mathcal{T}|$ is even.
Together, they suggest a highly efficient technique for finding cuts, the Cut-Search Algorithm (CSA) described in Algorithm~\ref{alg1}. If there is a violated parity inequality, the CSA returns the set $\mathcal{V}$ corresponding to the cut; otherwise, it returns an empty set.

As mentioned above, the CSA was used by Taghavi \emph{et al.}~\cite[Algorithm~2]{SparseLP} in conjunction with ALP decoding, and by Wadayama~\cite[Fig.~6]{Wadayama} as a feasibility check in the context of interior point decoding. In addition to providing another perspective on the CSA, the necessary condition and sufficient condition proved in Theorems 1 and 2, respectively, serve as the basis for a new adaptive approach to finding cut-inducing RPCs, as described in the next section.

\begin{algorithm}
\begin{doublespace}
\caption{Cut-Search Algorithm (CSA)}
\label{alg1}
\begin{algorithmic}[1]
\renewcommand{\algorithmicrequire}{\textbf{Input:}}
\renewcommand{\algorithmicensure}{\textbf{Output:}}
\REQUIRE parity-check node $j$ and vector $\mathbf u$
\ENSURE variable node set $\mathcal{V}$
\STATE $\mathcal{V}\leftarrow \mathcal{T} = \{i\in\mathcal{N}(j)|u_i>\frac{1}{2}\}$ and $\mathcal{S}\leftarrow \{i\in\mathcal{N}(j)|0<u_i<1\}$
\IF{$|\mathcal{V}|$ is even}
\IF {$\mathcal{S}\neq\varnothing$}
\STATE $i^*\leftarrow \arg\mathop {\min}\limits_{i\in\mathcal{S}} \left|\frac{1}{2}-u_i\right| $
\ELSE
\STATE $i^*\leftarrow$ arbitrary $i\in\mathcal{N}(j)$
\ENDIF
\IF {$i^*\in\mathcal{V}$}
\STATE $\mathcal{V}\leftarrow \mathcal{V}\setminus\{i^*\}$
\ELSE
\STATE $\mathcal{V}\leftarrow \mathcal{V}\cup\{i^*\}$
\ENDIF
\ENDIF
\IF {$\sum\limits_{i \in \mathcal{V}} \left(1-u_i\right)  + \sum\limits_{i \in \mathcal{N}(j)\backslash \mathcal{V}} {{u_i}}  < 1$}
\STATE Found the violated parity inequality on parity-check node $j$
\ELSE
\STATE There is no violated parity inequality on parity-check node $j$
\STATE $\mathcal{V}\leftarrow\emptyset$
\ENDIF
\RETURN $\mathcal{V}$
\end{algorithmic}
\end{doublespace}
\end{algorithm}

\section{LP Decoding with Adaptive Cut-Generating Algorithm}
\label{sec:ACGA}

\subsection{Generating Redundant Parity Checks}
\label{subsec:RPC}

Although the addition of a redundant row to a parity-check matrix does not affect the $\mathbb{F}_2$-nullspace and, therefore, the linear code it defines, different parity-check matrix representations of a linear code may give different fundamental polytopes underlying the corresponding LP relaxation of the ML decoding problem. This fact inspires the use of cutting-plane techniques to improve the error-correcting performance of the original LP and ALP decoders. Specifically, when the LP decoder gives a nonintegral solution (i.e., a pseudocodeword), we try to find the RPCs that introduce cuts at that point, if such RPCs exist. The cuts obtained in this manner are called \emph{RPC cuts}. The effectiveness of this method depends on how closely the new relaxation approximates the ML decoding problem, as well as on the efficiency of the technique used to search for the cut-inducing RPCs.

An RPC can be obtained by modulo-2 addition of some of the rows of the original parity-check matrix, and this new check introduces a number of linear constraints that may give a cut. In \cite{Taghavi_ALP}, a random walk on a cycle within the subgraph defined by the nonintegral entries in a pseudocodeword served as the basis for a search for RPC cuts.  However, there is no guarantee that this method will find a cut (if one exists) within a finite number of iterations. In fact, the average number of random trials needed to find an RPC cut grows exponentially with the code length.

The IP-based separation algorithm in \cite{SepAlg} performs Gaussian elimination on a submatrix comprising the columns of the original parity-check matrix that correspond to the nonintegral entries in a pseudocodeword in order to get redundant parity checks. In~\cite{cutting-plane}, the RPCs that potentially provide cutting planes are obtained by transforming a column-permuted version of the submatrix into row echelon form. The chosen permutation organizes the columns according to descending order of their associated nonintegral pseudocodeword entries, with the exception of the column corresponding to the largest nonintegral entry, which is placed in the rightmost position of the submatrix \cite[p.~1010]{cutting-plane}. This approach was motivated by the fact that a parity check $j$ provides a cut at a pseudocodeword if there exists a  variable node in $\mathcal{N}(j)$ whose value is greater than the sum of the values of all of the other neighboring variable nodes \cite[Lemma 2]{cutting-plane}. However, when combined with ALP decoding, the resulting ``cutting-plane algorithm'' does not provide sufficiently many cuts to surpass the separation algorithm in error-rate performance.

Motivated by the new derivation of the CSA based on the conditions in Theorems \ref{thm1} and \ref{thm2}, we next propose a new algorithm for generating cut-inducing RPCs.  When used with ALP decoding, the cuts have been found empirically to achieve near-ML decoding performance in the high-SNR region for several short-to-moderate length LDPC codes. However, application of these new techniques to codes with larger block lengths proved to be prohibitive computationally, indicating that further work is required to develop practical methods for enhanced LP decoding of longer codes.
%Although yet not tested on LDPC codes of large block length due to limited computational power, we believe that the proposed algorithm would also provide significant improvement in error-rate performance for longer codes.

Given a nonintegral solution of the LP problem, we can see from Theorems \ref{thm1} and \ref{thm2} that an RPC with a small number of nonintegral neighboring variable nodes may be more likely to satisfy the necessary condition for providing a cut at the pseudocodeword. Moreover, the nonintegral neighbors should have values either close to 0 or close to 1; in other words, they should be as far from $\frac{1}{2}$ as possible.

Let $\mathbf{p}=(p_1,p_2,\ldots,p_n)\in[0,1]^n$ be a pseudocodeword solution to LP decoding, with $a$ nonintegral positions, $b$ zeros, and $n-a-b$ ones. We first group entries of $\mathbf{p}$ according to whether their values are nonintegral, zero, or one. Then, we sort the nonintegral positions in ascending order according to the value of $\left|\frac{1}{2}-p_i\right|$ and define the permuted vector $\mathbf{p'}=\Pi(\mathbf{p})$ satisfying the following ordering
\begin{equation}
\label{Sortfrac}
\left|\frac{1}{2}-p'_1\right|\leq\dots\leq\left|\frac{1}{2}-p'_a\right|,
\end{equation}
\begin{equation}\notag
%\label{Sort0}
p'_{a+1}=\dots=p'_{a+b}=0,
\end{equation}
and%
\begin{equation}\notag
%\label{Sort1}
p'_{a+b+1}=\dots=p'_n=1.
\end{equation}
By applying the same permutation $\Pi$ to the columns of the original parity-check matrix $\mathbf{H}$, we get
\begin{equation}
\label{PiH}
\mathbf{H'} \triangleq \Pi(\mathbf{H}) = \left(\mathbf{H}^{(\mathrm{f})} |\mathbf{H}^{(0)}|\mathbf{H}^{(1)}\right)
\end{equation}
where $\mathbf{H}^{(\mathrm{f})}$, $\mathbf{H}^{(0)}$, and $\mathbf{H}^{(1)}$ consist of columns of $\mathbf H$ corresponding to positions of $\mathbf{p}'$ with nonintegral values, zeros, and ones, respectively.

The following familiar definition from matrix theory will be useful \cite[p. 10]{HornJohnson}.
\newtheorem{def3}[def1]{Definition}
\begin{def3}
\label{def_ref}
A matrix is in \emph{reduced row echelon form} if its  nonzero rows (i.e., rows with at least one nonzero element) are above any all-zero rows, and the leading entry (i.e., the first nonzero entry from the left) of a nonzero row is the only nonzero entry in its column and is always strictly to the right of the leading entry of the row above it.
\end{def3}

By applying a suitable sequence of elementary row operations $\Phi$ (over $\mathbb{F}_2$) to $\mathbf{H'}$, we get
\begin{equation}
\label{PhiH}
\mathbf{\bar H} \triangleq \Phi(\mathbf{H'}) = \left(\mathbf{\bar H}^{(\mathrm{f})} |\mathbf{\bar H}^{(0)}|\mathbf{\bar H}^{(1)}\right),
\end{equation}
where  $\mathbf{\bar H}^{(\mathrm{f})}$ is in reduced row echelon form. Applying the inverse permutation $\Pi^{-1}$ to columns of $\mathbf{\bar H}$, we get an equivalent parity-check matrix \begin{equation}
\label{tildeH}
\mathbf{\tilde H}=\Pi^{-1}(\mathbf{\bar H})
\end{equation}
whose rows are likely to be cut-inducing RPCs, for the reasons stated above.

Multiple nonintegral positions in the pseudocodeword $\mathbf{p}$ could have values of the same distance from $\frac{1}{2}$, i.e., $|\frac{1}{2}-p_i|=|\frac{1}{2}-p_j|$ for some $i\neq j$.  In such a case, the ordering of the nonintegral positions in \eqref{Sortfrac} is not uniquely determined. Hence, the set of RPCs generated by operations \eqref{PiH}--\eqref{tildeH} may depend upon the particular ordering reflected in the permutation $\Pi$. Nevertheless, if the decoder uses a fixed, deterministic sorting rule such as, for example, a stable sorting algorithm, then the decoding error probability will be independent of the transmitted codeword.

%Note that, if more than one nonintegral positions in pseudocodeword $\mathbf{p}$ have the same distance to $\frac{1}{2}$, i.e., $|\frac{1}{2}-p_i|=|\frac{1}{2}-p_j|$ for some $i\neq j$, there can be more than one ascending ordering for these nonintegral positions. Hence, the set of RPCs generated by the above operations \eqref{PiH}--\eqref{tildeH} can depend on how the permutation $\Pi$ sorts the nonintegral positions. Nevertheless, if the ordering of such positions is always done in a consistent manner, or a stable sorting algorithm is used, the decoding error probability is independent of the transmitted codeword.

The next theorem describes a situation in which a row of $\mathbf{\tilde H}$ is guaranteed to provide a cut.
\newtheorem{thm3}[thm1]{Theorem}
\begin{thm3}
\label{thm3}
If there exists a weight-one row in submatrix $\mathbf{\bar H}^{(\mathrm{f})}$, the corresponding row of the equivalent parity-check matrix $\mathbf{\tilde H}$ is a cut-inducing RPC.
\end{thm3}
\begin{IEEEproof}
Given a pseudocodeword $\mathbf{p}$, suppose the $j$th row of submatrix $\mathbf{\bar H}^{(\mathrm{f})}$ has  weight one and the corresponding nonintegral position in $\mathbf{p}$ is $p_i$. Since it is the only nonintegral position in $\mathcal{N}(j)$, the left-hand side of \eqref{CUT_SCe} is equal to $-\left|\frac{1}{2}-p_i\right|$. Since $0 < p_i < 1$, this is larger than $-\frac{1}{2}$, the right-hand side. Hence, according to Theorem \ref{thm2}, RPC $j$ satisfies the sufficient condition for providing a cut. In other words, there must be a violated parity inequality induced by RPC $j$.
\end{IEEEproof}

\newtheorem{thm3rm}[thm1rm]{Remark}
\begin{thm3rm}
\label{thm3rm}
Theorem~\ref{thm3} is equivalent to \cite[Theorem~3.3]{SepAlg}. The proof of the result shown here, though, is considerably simpler, thanks to the application of Theorem~\ref{thm2}.
\end{thm3rm}

Although Theorem~\ref{thm3} only ensures a cut for rows with weight one in submatrix $\mathbf{\bar H}^{(\mathrm{f})}$, rows in $\mathbf{\bar H}^{(\mathrm{f})}$ of weight larger than one may also provide RPC cuts. Hence, the CSA should be applied on every row of the redundant parity-check matrix $\mathbf{\tilde H}$ to search for all possible RPC cuts.
The approach of generating a redundant parity-check matrix $\mathbf{\tilde H}$ based on a given pseudocodeword and applying the CSA on each row of this matrix is called adaptive cut generation (ACG). Combining  ACG with ALP decoding, we obtain the ACG-ALP decoding algorithm described in Algorithm~\ref{alg2}.
Beginning with the original parity-check matrix, the algorithm iteratively applies ALP decoding. When a point is reached when no further cuts can be produced from the original parity-check matrix, the ACG technique is invoked to see whether any RPC cuts can be generated. The ACG-ALP decoding iteration stops when no more cuts can be found either from the original parity-check matrix or in the form of redundant parity checks.

\begin{algorithm}
\begin{doublespace}
\caption{Adaptive Linear Programming with Adaptive Cut-Generation (ACG-ALP) Decoding Algorithm}
\label{alg2}
\begin{algorithmic}[1]
\renewcommand{\algorithmicrequire}{\textbf{Input:}}
\renewcommand{\algorithmicensure}{\textbf{Output:}}
\REQUIRE cost vector $\boldsymbol{\gamma}$, original parity-check matrix $\mathbf{H}$
\ENSURE Optimal solution of current LP problem
\STATE Initialize the LP problem with the constraints in \eqref{init_cons}.
\STATE Solve the current LP problem, and get optimal solution $x^*$.
\STATE Apply {\bf Algorithm 1 (CSA)} on each row of $\mathbf{H}$.
%\IF{No cut is found \AND $x^*$ is nonintegral}
\IF{No cut is found {\bf and} $x^*$ is nonintegral}
\STATE Construct $\mathbf{\tilde H}$ associated with $x^*$ according to \eqref{PiH}--\eqref{tildeH}.
\STATE Apply {\bf Algorithm 1 (CSA)} to each row of $\mathbf{\tilde H}$.
\ENDIF
\IF {No cut is found}
\STATE Terminate.
\ELSE
\STATE Add cuts that are found into the LP problem as constraints, and go to line 2.
\ENDIF
\end{algorithmic}
\end{doublespace}
\end{algorithm}

\subsection{Reducing the Number of Constraints in the LP Problem}
\label{subsec:malp}

In the ALP decoding, the number of constraints in the LP problem grows as the number of iterations grows, increasing the complexity of solving the LP problem. For ACG-ALP decoding, this problem becomes more severe since the algorithm generates additional RPC cuts and uses more iterations to successfully decode inputs on which the ALP decoder has failed.

From Lemma~\ref{lemma1}, we know that a binary parity-check constraint can provide at most one cut. Hence, if a binary parity check gives a cut, all other linear inequalities introduced by this parity check in previous iterations can be removed from the LP problem. The implementation of this observation leads to a \emph{modified ALP (MALP)} decoder referred to as the MALP-A decoder~\cite{SparseLP}.
This decoder improves the efficiency of ALP decoding, where only cuts associated with the original parity-check matrix are used. However, with ACG-ALP decoding, different RPCs may be generated adaptively in every iteration and most of them give only one cut throughout the sequence of decoding iterations. As a result, when MALP-A decoding is combined with the ACG technique, only a small number of constraints are removed from the LP problem, and the decoding complexity is only slightly reduced.

\newtheorem{def4}[def1]{Definition}
\begin{def4}
\label{def_act}
A linear inequality constraint of the form $\mathbf{ax}\geq b$ is called \emph{active} at point $\mathbf{x}^*$ if it holds with equality, i.e., $\mathbf{ax}^*= b$, and is called \emph{inactive} otherwise.
%if it holds with strict inequality, i.e., $\mathbf{ax}^* > b$.
\end{def4}

For an LP problem with a set of linear inequality constraints, the optimal solution $\mathbf{x}^*\in[0,1]^n$ is a vertex of the polytope formed by the hyperplanes corresponding to all active constraints. In other words, if we set up an LP problem with only those active constraints, the optimal solution remains the same. Therefore, a simple and intuitive way to reduce the number of constraints is to remove all inactive constraints from the LP problem at the end of each iteration, regardless of whether or not the corresponding binary parity check generates a cut. This approach is called MALP-B decoding~\cite{SparseLP}. By combining the ACG technique and the MALP-B algorithm, we obtain the ACG-MALP-B decoding algorithm. It is similar to the ACG-ALP algorithm described in Algorithm~\ref{alg2} but includes one additional step that removes all inactive constraints from the LP problem, as indicated in Line 3 of Algorithm~\ref{alg3}.

Since adding further constraints into an LP problem reduces the feasible space, the minimum value of the cost function is non-decreasing as a function of the number of iterations. In our computer simulations, the ACG-MALP-B decoding algorithm was terminated when no further cuts could be found.
(See Fig.~\ref{fig_iter} for statistics on the average number of iterations required to decode one codeword of the (155,64) Tanner LDPC code.)

%Since adding further constraints into an LP problem reduces the feasible space, the minimum value of the cost function in each iteration does not decrease.
%The modified decoding algorithm will terminate when no further cuts can be found. However, it is theoretically possible that there exist more than one optimal vertices of the fundamental polytope that give the same optimal value of the cost function. In such situation, after cutting off one of the optimal vertices, say $V_1$, by adding extra constraints, the solution of the underlying LP problem in next iteration may become another optimal vertex, say $V_2$, that still gives the same optimal value, while the added constraints may or may not be active depending on the types of LP solvers used. If the extra constraints added at the end of previous iteration are not active, then they will be removed at the end of current iteration. Hence, in the next iteration after adding constraints that cut off $V_2$, the optimal point of the underlying LP problem may become $V_1$ again. Therefore, it is possible that the optimal points of the LP problems in following iterations alternate between $V_1$ and $V_2$. To prevent the decoding algorithm from endless iterations, we can simply set a maximum number of allowed iterations. We remark that, in our own application of the ACG-MALP-B decoding algorithm, we have never seen an instance of such situation, and the algorithm has only stopped when no further cuts could be found.

\begin{algorithm}[t]
\begin{doublespace}
\caption{ACG-MALP-B/C Decoding Algorithm}
\label{alg3}
\begin{algorithmic}[1]
\renewcommand{\algorithmicrequire}{\textbf{Input:}}
\renewcommand{\algorithmicensure}{\textbf{Output:}}
\REQUIRE cost vector $\boldsymbol{\gamma}$, original parity-check matrix $\mathbf{H}$
\ENSURE Optimal solution of current LP problem
\STATE Initialize LP problem with the constraints in \eqref{init_cons}.
\STATE Solve the current LP problem, get optimal solution $x^*$.
\STATE ACG-MALP-B only: remove all inactive constraints from the LP problem.
\STATE ACG-MALP-C only: remove inactive constraints that have above-average slack values from the LP problem.
\STATE Apply {\bf CSA} only on rows of $\mathbf{H}$ that have not introduced constraints.
%\IF{No cut is found \AND $x^*$ is nonintegral}
\IF{No cut is found {\bf and} $x^*$ is nonintegral}
\STATE Construct $\mathbf{\tilde H}$ according to $x^*$
\STATE Apply {\bf CSA} on each row of $\mathbf{\tilde H}$.
\ENDIF
\IF {No cut is found}
\STATE Terminate.
\ELSE
\STATE Add found cuts into LP problem as constraints, and go to line 2.
\ENDIF
\end{algorithmic}
\end{doublespace}
\end{algorithm}

In our implementation of both MALP-B and ACG-MALP-B decoding, we have noticed that a considerable number of the constraints deleted in previous iterations are added back into the LP problem in later iterations, and, in fact,  many of them are added and deleted several times. We have observed that MALP-B-based decoding generally takes more iterations to decode a codeword than ALP-based decoding, resulting in a tradeoff between the number of iterations and the size of the constituent LP problems. MALP-B-based decoding has the largest number of iterations and the smallest LP problems to solve in each iteration, while ALP-based decoding  has a smaller number of iterations but larger LP problems.

Although it is difficult to know in advance which inactive constraints might become cuts in later iterations, there are several ways to find a better tradeoff between the MALP-B and ALP techniques to speed up LP decoding. This tradeoff, however, is highly dependent on the LP solver used in the implementation. For example, we used the Simplex solver from the open-source GNU Linear Programming Kit (GLPK)~\cite{glpk}, and found that the efficiency of iterative ALP-based decoders is closely related to the total number of constraints used to decode one codeword, i.e., the sum of the number of constraints used in all iterations. This suggests a new criterion for the removal of inactive constraints whose implementation we call the MALP-C decoder.

In MALP-C decoding, instead of removing all inactive constraints from the LP problem in each iteration, we remove only the linear inequality constraints with slack variables that have above-average values, as indicated in Line 4 of Algorithm~\ref{alg3}. The ACG-MALP-B and ACG-MALP-C decoding algorithms are both described in Algorithm~\ref{alg3}, differing only in the use of Line 3 or Line 4. Although all three of the adaptive variations of LP decoding discussed in this paper -- ALP, MALP-B, and MALP-C --  have the exact same error-rate performance as the original LP decoder, they may lead to different decoding results for a given received vector when combined with the ACG technique, as shown in the next section.

\section{Numerical Results}
\label{sec:NR}

To demonstrate the  improvement offered by our proposed decoding algorithms, we compared their error-correcting performance to that of ALP decoding (which, again, has the same performance as the original LP decoding), BP decoding (two cases, using the sum-product algorithm with a maximum of 100 iterations and 1000 iterations, respectively), the separation algorithm (SA) \cite{SepAlg}, the random-walk-based RPC search algorithm \cite{Taghavi_ALP}, and  ML decoding for various LDPC codes on the additive white Gaussian noise (AWGN) channel. We use the Simplex algorithm from the open-source GLPK \cite{glpk} as our LP solver. The LDPC codes we evaluated are MacKay's rate-$\frac{1}{2}$, (3,6)-regular LDPC codes with lengths 96 and 408, respectively \cite{MackayCode}; a  rate-$\frac{1}{4}$,  (3,4)-regular LDPC code of length 100; the rate-$\frac{2}{5}$, (3,5)-regular Tanner code of length 155 \cite{Tanner}; and a rate-0.89, (3,27)-regular high-rate LDPC code of length 999 \cite{MackayCode}.

The proposed ACG-ALP, ACG-MALP-B, and ACG-MALP-C decoding algorithms are all based on the underlying cut-searching algorithm (Algorithm~\ref{alg1}) and the adaptive cut-generation technique of Section~\ref{subsec:RPC}. Therefore, their error-rate performance is very similar. However, their performance may not be identical, because cuts are found adaptively from the output pseudocodewords in each iteration and the different sets of constraints used in the three proposed algorithms may lead to different solutions of the corresponding LP problems.

In our simulation, the LP solver uses double-precision floating-point arithmetic, and therefore, due to this limited numerical resolution, it may round some small nonzero vector coordinate values to 0 or output small nonzero values for vector coordinates which should be 0. Similar rounding errors may occur for coordinate values close to 1. Coordinates whose values get rounded to integers by the LP solver might lead to some ``false'' cuts -- parity inequalities not actually violated by the exact LP solution.  This is because such rounding by the LP solver would decrease the left-hand side of parity inequality \eqref{PI1}. On the other hand, when coordinates that should have integer values are given nonintegral values, the resulting errors would increase the left-hand side of parity inequality \eqref{PI1}, causing some cuts to be missed.  Moreover, this would also increase the size of the submatrix $\mathbf{H}^{(\mathrm{f})}$ in \eqref{PiH},  leading to higher complexity for the ACG-ALP decoding algorithm.

To avoid such numerical problems in our implementation of the CSA, we used $1-10^{-6}$ instead of 1 on the right-hand side of the inequality in line 14 of Algorithm~\ref{alg1}. Whenever the LP solver outputs a solution vector, coordinates with value less than $10^{-6}$ were rounded to 0 and coordinates with value larger than $1-10^{-6}$ were rounded to 1. The rounded values were then used in the cut-search and RPC-generation steps in the decoding algorithms described in previous sections. If such a procedure were not applied, and if, as a result, false cuts were to be produced, the corresponding constraints, when added into the LP problem to be solved in the next step, would leave the solution vector unchanged, causing the decoder to become stuck in an endless loop. We saw no such behavior in our decoder simulations incorporating the prescribed thresholding operations.

Finally, we want to point out that there exist LP solvers, such as  \emph{QSopt\_ex Rational LP Solver} \cite{QSopt}, that produce exact rational solutions to LP instances with  rational input.  However, such solvers generally have higher computational overhead than their floating-point counterparts. For this reason, we did not use an exact rational LP solver in our empirical studies.

\begin{figure}[!t]
\includegraphics[width=0.9\linewidth]{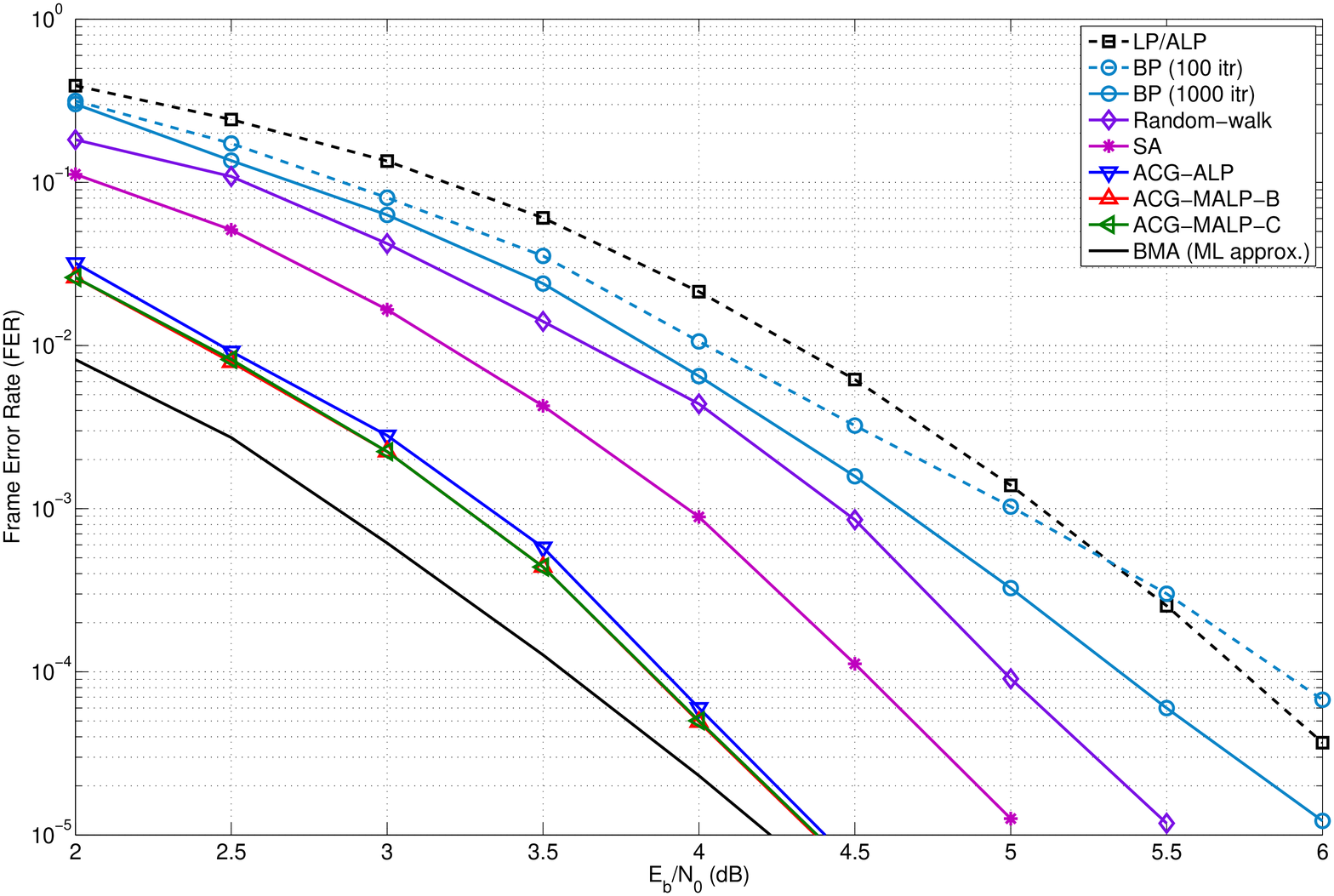}
\centering
\caption{FER versus $E_b/N_0$ for random (3,4)-regular LDPC code of length 100 on the AWGN channel.}\label{fig_fer_100}
\end{figure}

Fig.~\ref{fig_fer_100} shows the simulation results for the length-100, regular-(3,4) LDPC code whose FER performance was also evalutated in \cite{Taghavi_ALP} and \cite{SepAlg}.
We can see that the proposed algorithms have a gain of about 2~dB over the original LP and ALP decoder. They also perform significantly better than both the separation algorithm and the random-walk algorithm. The figure also shows
the results obtained with the Box-and-Match soft-decision decoding algorithm (BMA) \cite{BMA}, whose FER performance is guaranteed to be within a factor of 1.05 times that of ML decoding. We conclude that the performance gap between the proposed decoders and ML decoding is less than 0.2~dB at an FER of $10^{-5}$.

\begin{figure*}[!t]
\includegraphics[width=0.9\linewidth]{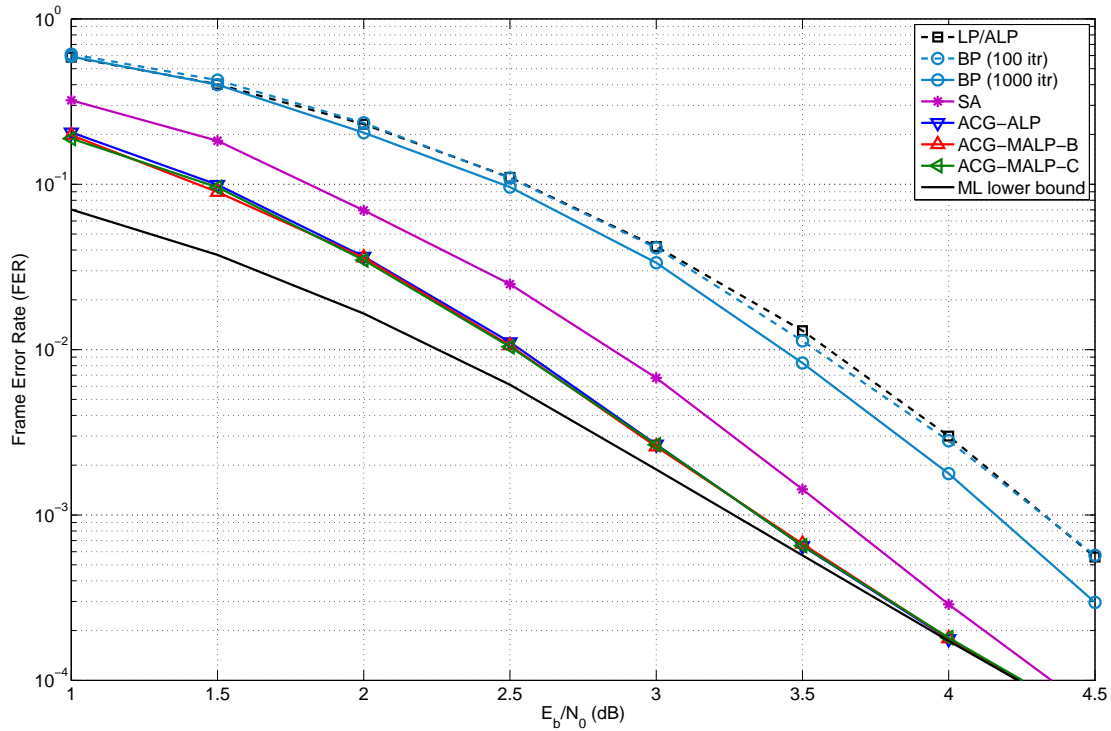}
\centering
\caption{FER versus $E_b/N_0$ for MacKay's random (3,6)-regular LDPC code of length 96 on the AWGN channel.}\label{fig_fer_96}
\end{figure*}

\begin{table*}
\renewcommand{\arraystretch}{1.3}
\caption{Frame Errors of ACG-ALP decoder on MacKay's random (3,6)-regular LDPC code of length 96 on the AWGN channel.}
\label{tabMLLB}
\centering
\begin{tabular}{|c|c|c|c|c|}
\hline
\bfseries $\text{E}_{\text{b}}/\text{N}_0$ (dB) & \bfseries Transmitted Frames & \bfseries Error Frames & \bfseries Pseudocodewords & \bfseries Incorrect Codewords\\
\hline
%3.0 & 386,445 & 1,000 & 255 & 745\\
%3.5 & 1,493,801 & 1,000 & 126 & 874\\
%4.0 & 5,589,843 & 1,000 & 23 & 977\\
%4.5 & 19,128,086 & 1,000 & 4 & 996\\
%5.0 & 61,198,400 & 1,000 & 0 & 1,000\\
%5.5 & 225,764,513 & 1,000 & 0 & 1,000\\
3.0 & 1,136,597 & 3,000 & 857 & 2,143\\
3.5 & 4,569,667 & 3,000 & 395 & 2,605\\
4.0 & 16,724,921 & 3,000 & 103 & 2,897\\
4.5 & 54,952,664 & 3,000 & 12 & 2,988\\
5.0 & 185,366,246 & 3,000 & 0 & 3,000\\
5.5 & 665,851,530 & 3,000 & 0 & 3,000\\
\hline
\end{tabular}
\end{table*}

In Fig.~\ref{fig_fer_96}, we show simulation results for MacKay's length-96, (3,6)-regular LDPC code (the 96.33.964 code from  \cite{MackayCode}). Again, the proposed ALP-based decoders with ACG demonstrate superior performance to the original LP, BP, and SA decoders over the range of SNRs considered. Table~\ref{tabMLLB} shows the actual frame error counts for the ACG-ALP decoder, with frame errors classified as either pseudocodewords or incorrect codewords; the ACG-MALP-B and ACG-MALP-C decoder simulations yielded very similar results. We used these counts to obtain a lower bound on ML decoder performance, also shown in the figure, by dividing the number of times the ACG-ALP decoder converged to an incorrect codeword by the total number of frames transmitted. Since the ML certificate property of LP decoding implies that ML decoding would have produced the same incorrect codeword in all of these instances, this ratio represents a lower bound on the FER of the ML decoder. We note that, when $E_b/N_0$ is greater than 4.5~dB, all decoding errors correspond to incorrect codewords, indicating that the ACG-ALP decoder has achieved ML decoding performance for the transmitted frames.

\begin{figure}
\includegraphics[width=0.9\linewidth]{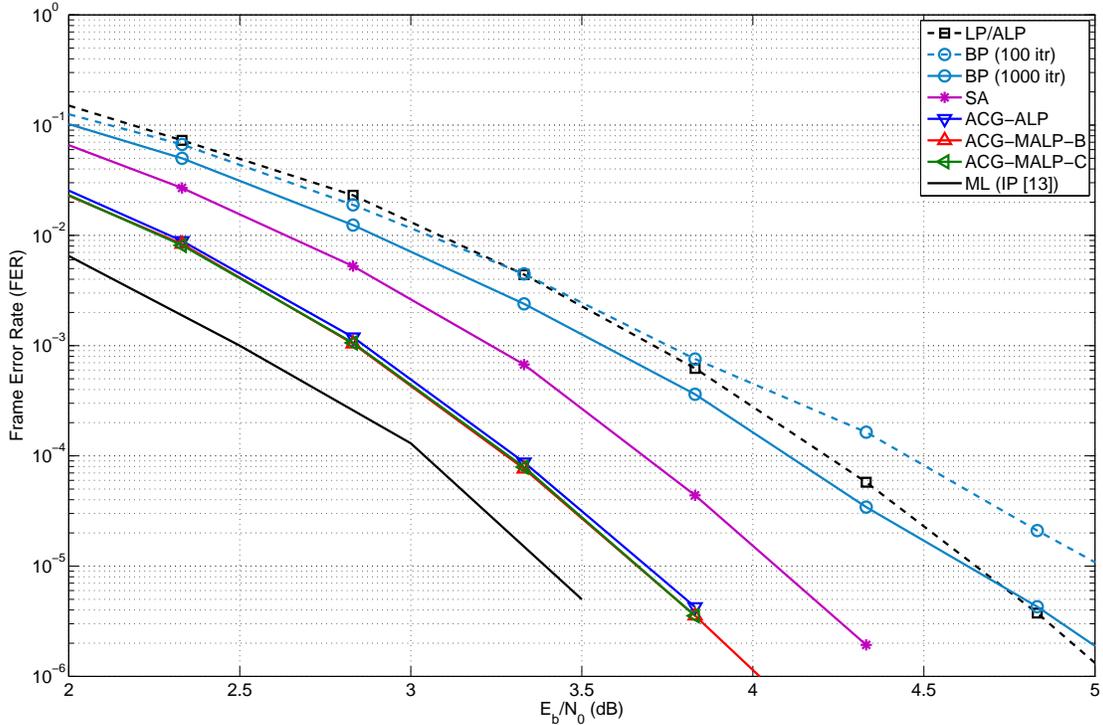}
\centering
\caption{FER versus $E_b/N_0$ for (155,64) Tanner LDPC code on the AWGN channel.}\label{fig_fer_155}
\end{figure}

Fig.~\ref{fig_fer_155} compares the performance of several different decoders applied to the (3,5)-regular, (155,64) Tanner code, as well as the ML performance curve from \cite{SepAlg}. It can be seen that the proposed ACG-ALP-based algorithms narrow the 1.25~dB gap between the original LP decoding and ML decoding to approximately 0.25~dB.

\begin{figure}
\includegraphics[width=0.9\linewidth]{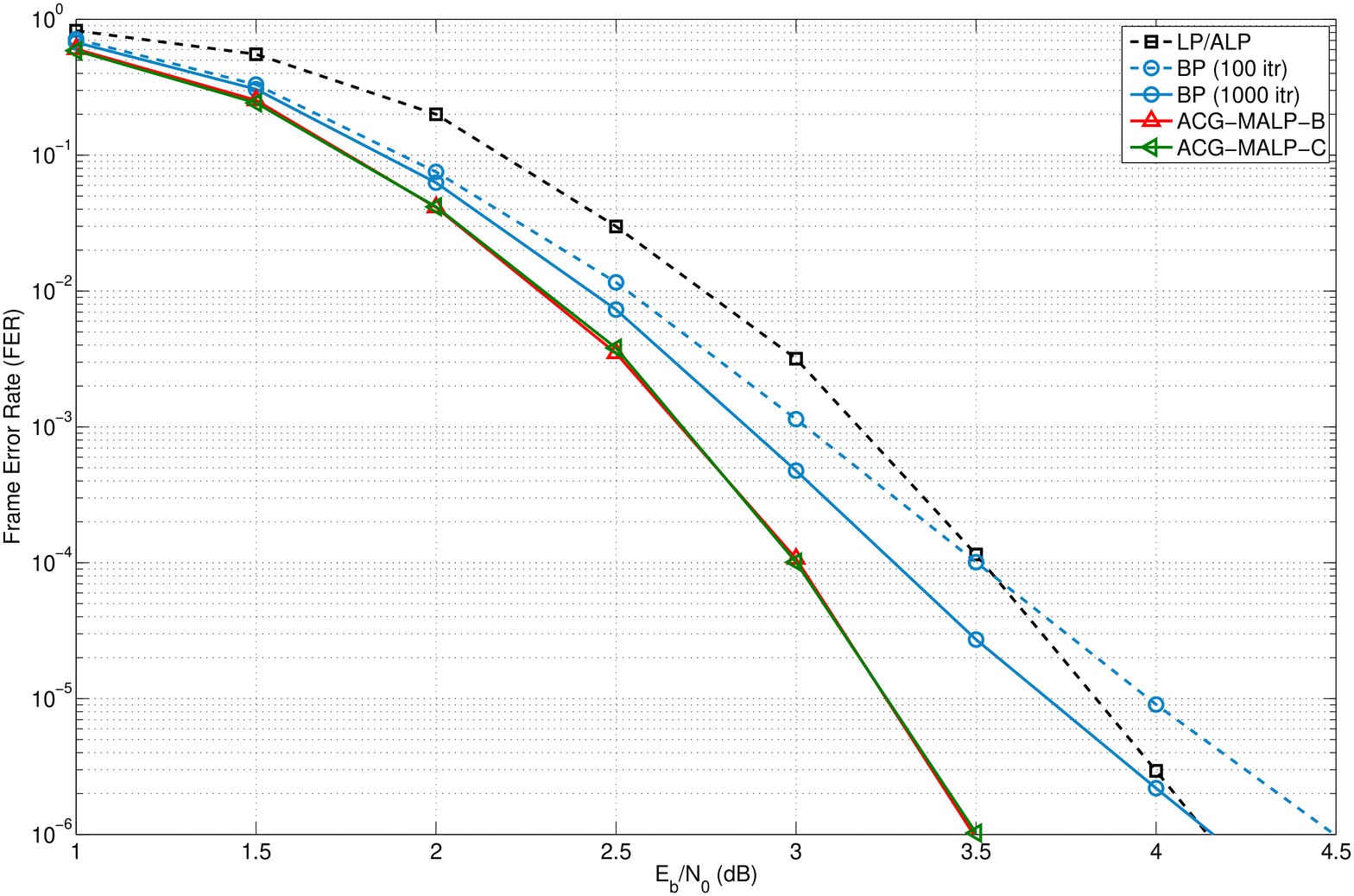}
\centering
\caption{FER versus $E_b/N_0$ for MacKay's random (3,6)-regular LDPC code of length 408 on the AWGN channel.}\label{fig_fer_408}
\end{figure}

\begin{figure}
\includegraphics[width=0.9\linewidth]{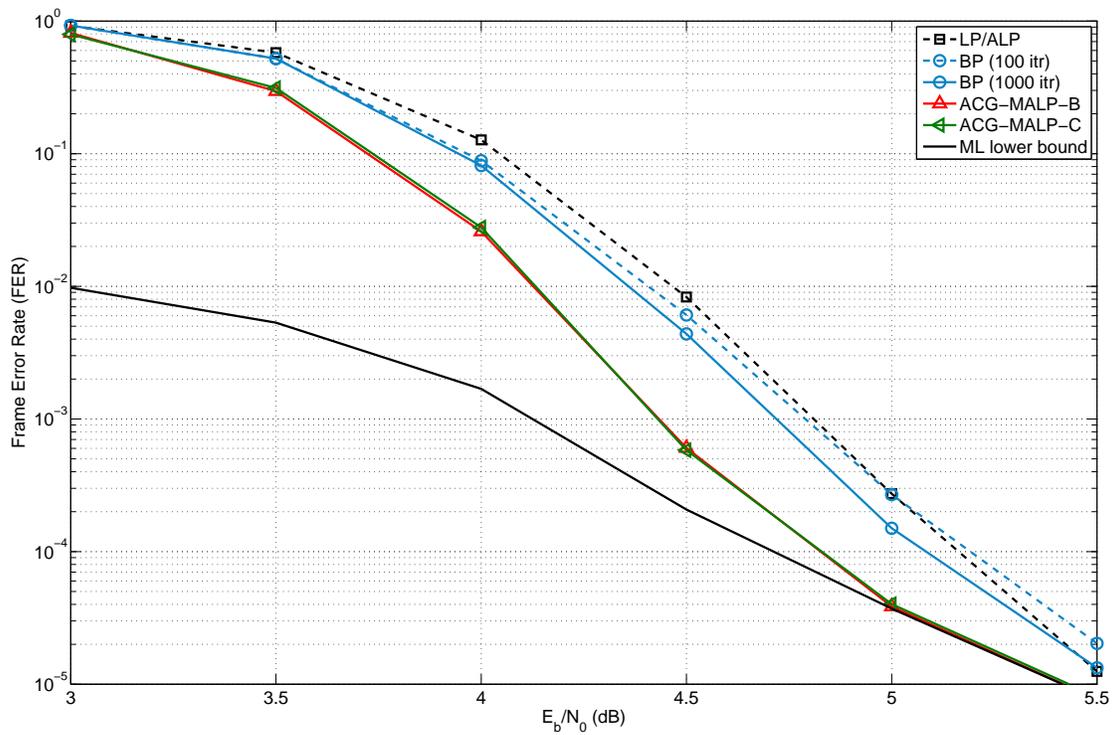}
\centering
\caption{FER versus $E_b/N_0$ for MacKay's random (3,27)-regular LDPC code of length 999 on the AWGN channel.}\label{fig_fer_999}
\end{figure}

We also considered two longer codes, MacKay's rate-$\frac{1}{2}$, random (3,6)-regular LDPC code of length 408 (the 408.33.844 code from \cite{MackayCode}) and a rate-0.89 LDPC code of length 999 (the 999.111.3.5543 code from \cite{MackayCode}). Because of the increased complexity of the constituent LP problems, we only simulated the ACG-MALP-B and ACG-MALP-C decoders.
In Fig.~\ref{fig_fer_408}, it is confirmed that the proposed decoding algorithms provide significant gain over the original LP decoder and the BP decoder, especially in the high-SNR region.
The results for the high-rate LDPC code, as shown in Fig.~\ref{fig_fer_999}, again show that the proposed decoding algorithms approaches ML decoding performance for some codes, where the ML lower bound is obtained using the same technique as in Fig.~\ref{fig_fer_96}. However, for the code of length 408, we found that the majority of decoding failures corresponded to pseudocodewords, so, in constrast to  the case of the length-96 and length-999 MacKay codes discussed above, the frame error data do not provide a good lower bound on ML decoder performance to use as a benchmark.

Since the observed improvements in ACG-ALP-based decoder performance comes from the additional RPC cuts found in each iteration, these decoding algorithms generally require more iterations and/or the solution of larger LP problems in comparison to ALP decoding. In the remainder of this section, we empirically investigate the relative complexity of our proposed algorithms in terms of such statistics as the average number of iterations, the average size of constituent LP problems, and the average number of cuts found in each iteration. All statistical data presented here were obtained from simulations of the Tanner (155,64) code on the AWGN channel. We ran all simulations until at least 200 frame errors were counted.

\begin{figure}[!t]
\includegraphics[width=0.9\linewidth]{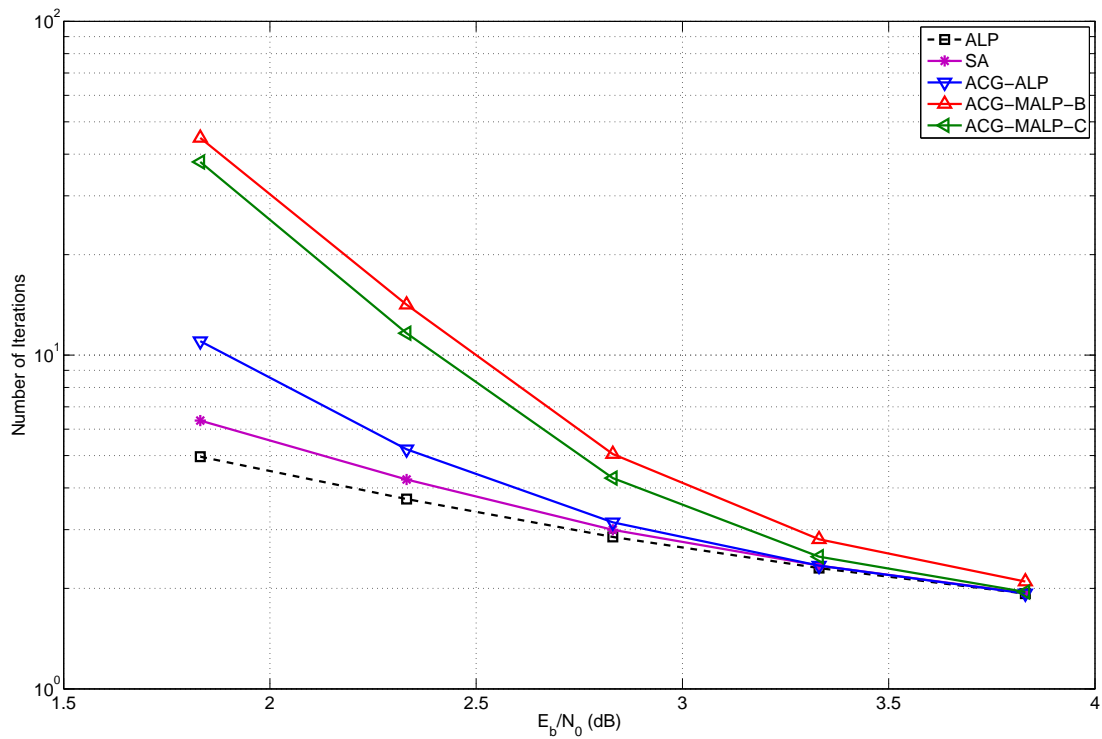}
\centering
\caption{Average number of iterations for decoding one codeword of (155,64) Tanner LDPC code.}\label{fig_iter}
\end{figure}

\begin{figure}[!t]
\centerline
{\subfigure[Average number of constraints in final iteration]{\includegraphics[width=0.45\linewidth]{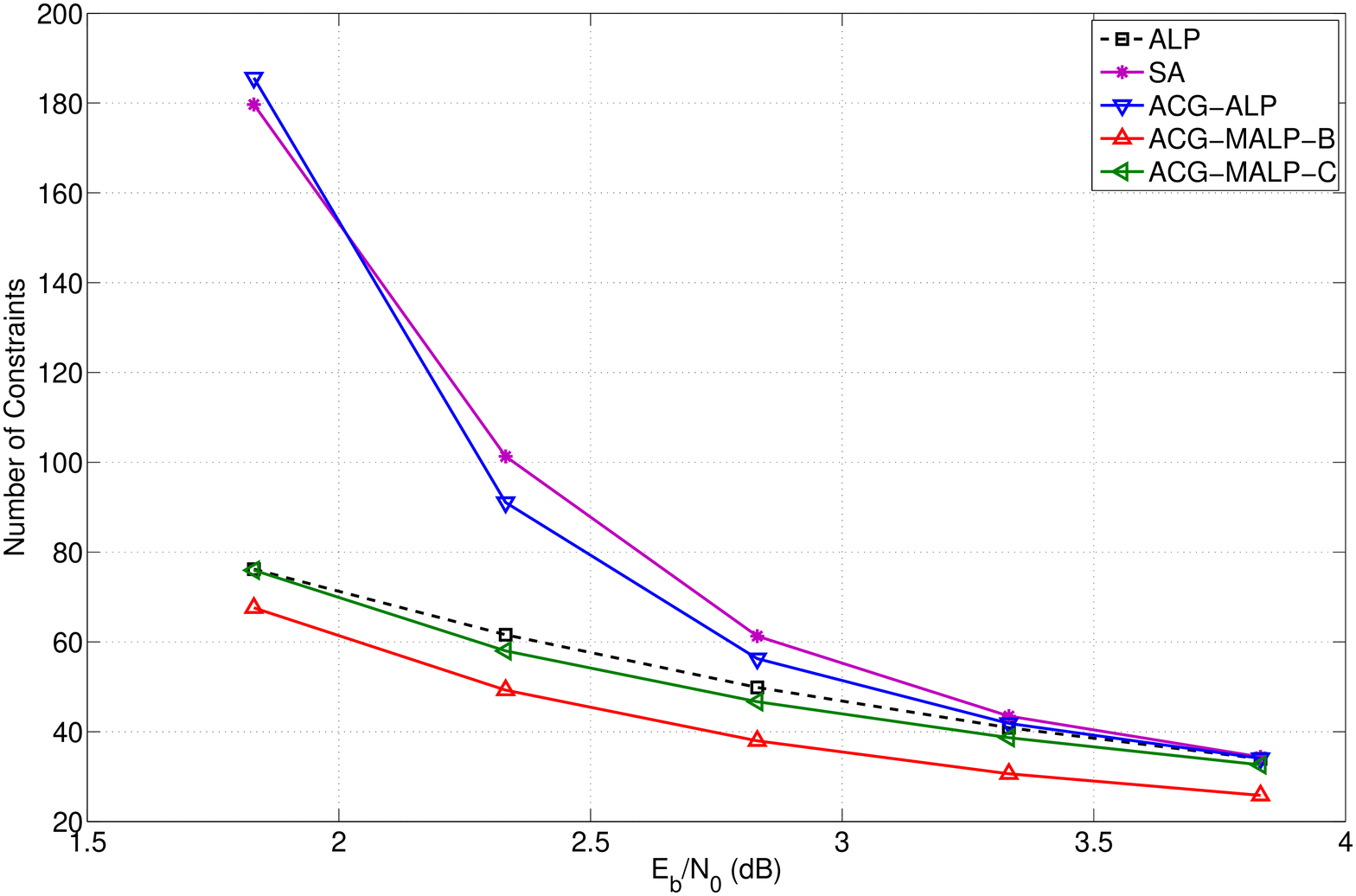}
\label{subfig_cst}}
\hfil
\subfigure[Average number of cuts found per iteration]{\includegraphics[width=0.45\linewidth]{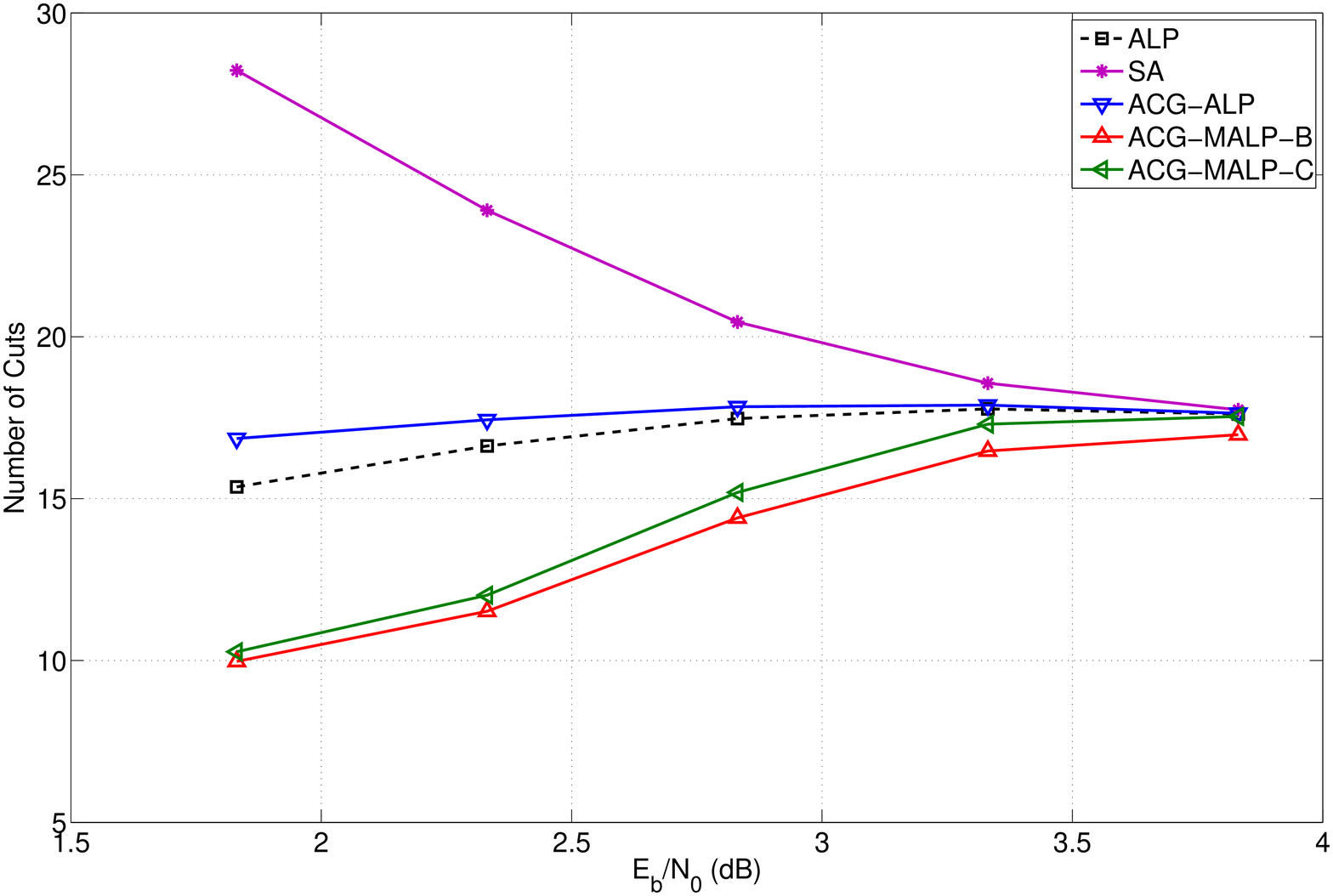}
\label{subfig_CpI}}}
\centerline
{\subfigure[Average number of cuts found from $\mathbf H$ for decoding one codeword]{\includegraphics[width=0.45\linewidth]{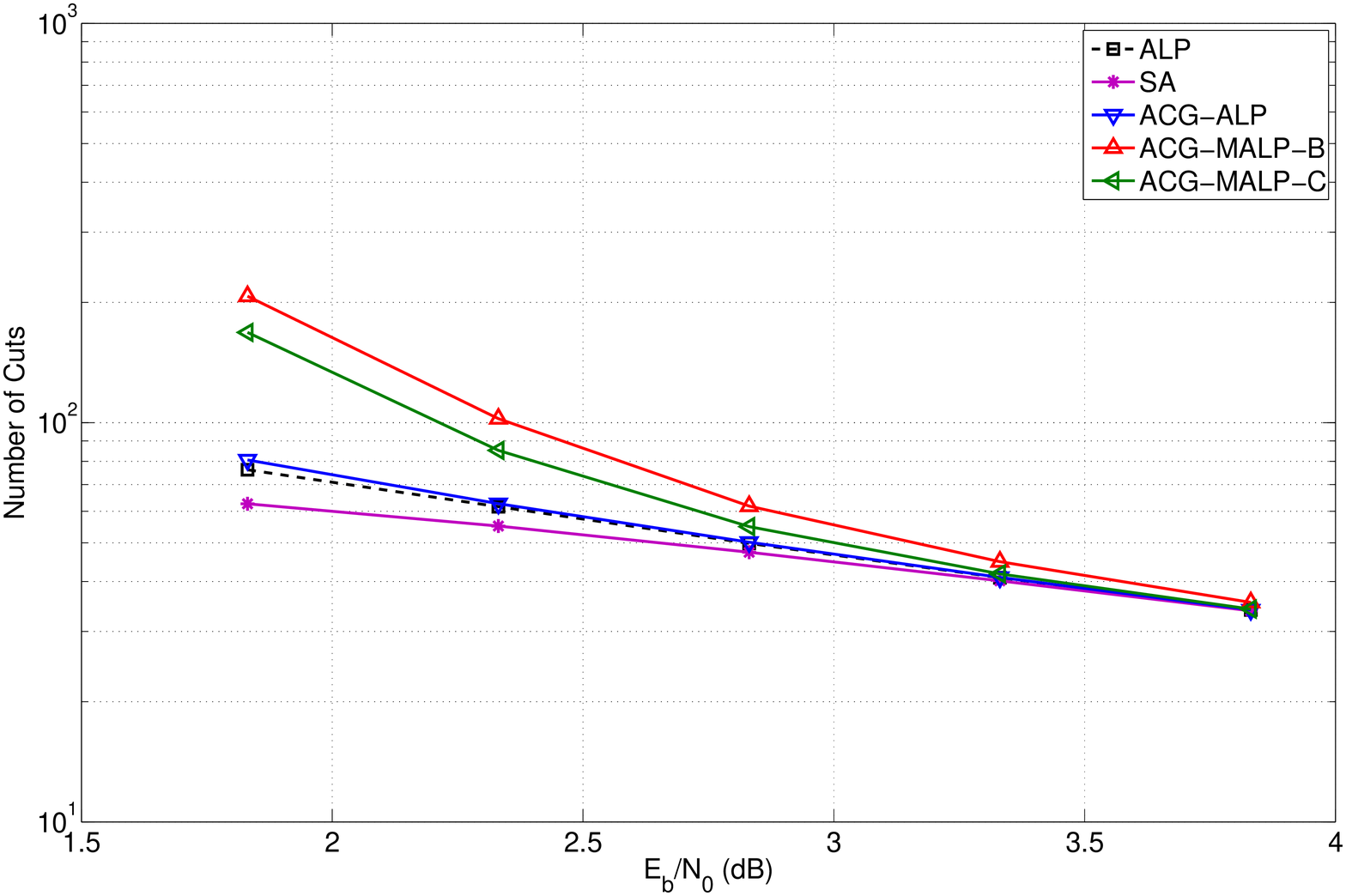}
\label{subfig_Hcut}}
\hfil
\subfigure[Average number of cuts found from RPCs for decoding one codeword]{\includegraphics[width=0.45\linewidth]{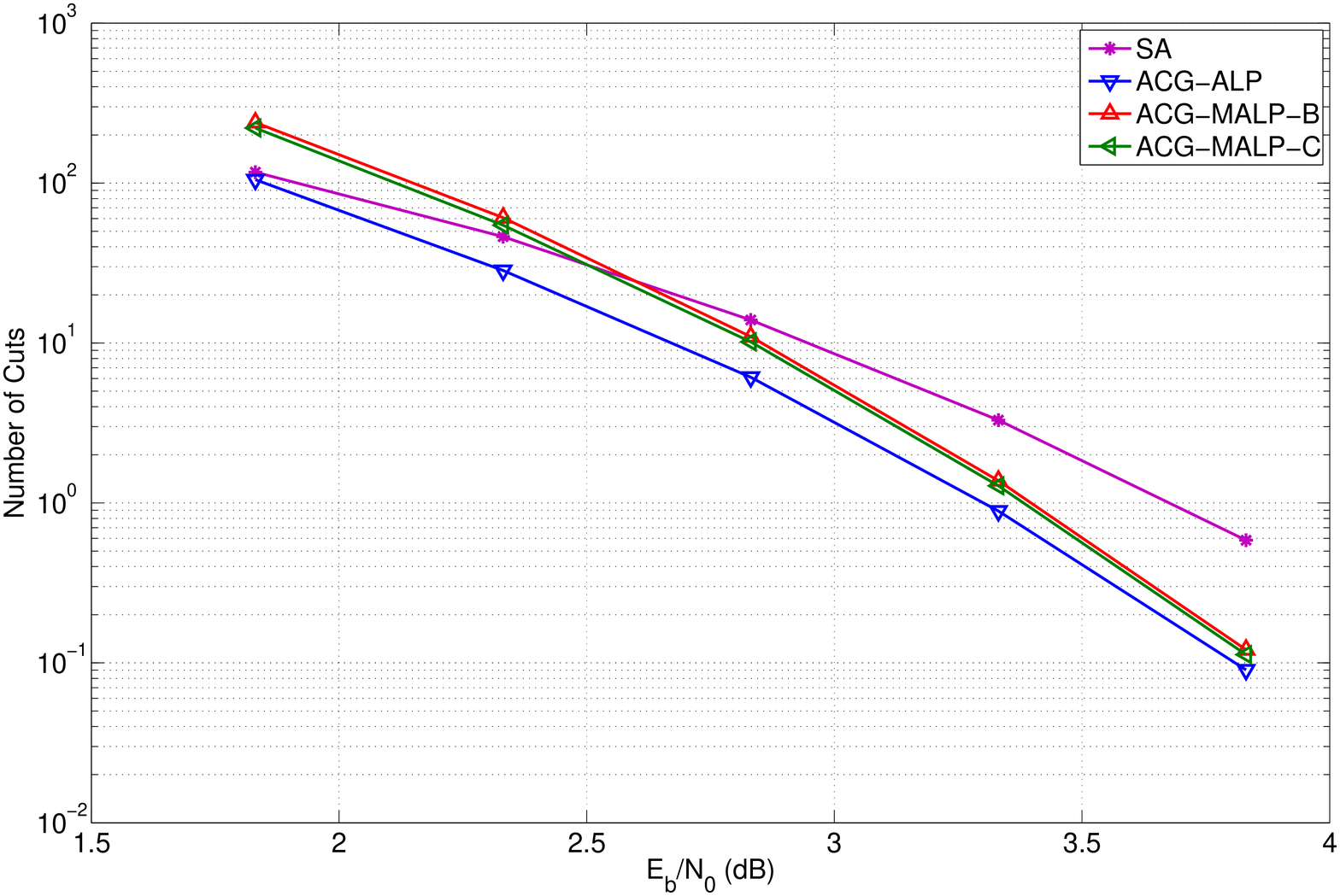}
\label{subfig_Rcut}}}
\caption{Average number of constraints/cuts during decoding iterations for decoding one frame of (155,64) Tanner LDPC code.}
\label{fig_cuts}
\end{figure}

In Fig.~\ref{fig_iter}, we compare the average number of iterations needed, i.e., the average number of LP problems solved, to decode one codeword. Fig.~\ref{subfig_cst} compares the average number of constraints in the LP problem of the final iteration that results in either a valid codeword or a pseudocodeword with no more cuts to be found. In Fig.~\ref{subfig_CpI}, we show the average number of cuts found and added into the LP problem in each iteration. Fig.~\ref{subfig_Hcut} and Fig.~\ref{subfig_Rcut} show the average number of cuts found from the original parity-check matrix $\mathbf H$ and from the generated RPCs, respectively.

\begin{table*}
\renewcommand{\arraystretch}{1.3}
\caption{The average accumulated number of constraints in all iterations of decoding one codeword of (155,64) Tanner code on the AWGN channel}
\label{tabCST}
\centering
\begin{tabular}{|c|c|c|c|c|}
\hline
\bfseries $\text{E}_{\text{b}}/\text{N}_0$ (dB) & \bfseries ACG-ALP & \bfseries ACG-MALP-B & \bfseries ACG-MALP-C\\
\hline
1.83 & 5495.8 & 5223.3 & 4643.1\\
2.33 & 1401.2 & 1387.3 & 1217.0\\
2.83 & 339.7 & 326.9 & 300.9\\
3.33 & 111.0 & 106.4 & 105.4\\
3.83 & 64.3 & 58.8 & 62.8\\
\hline
\end{tabular}
\end{table*}

\begin{figure}
\includegraphics[width=0.9\linewidth]{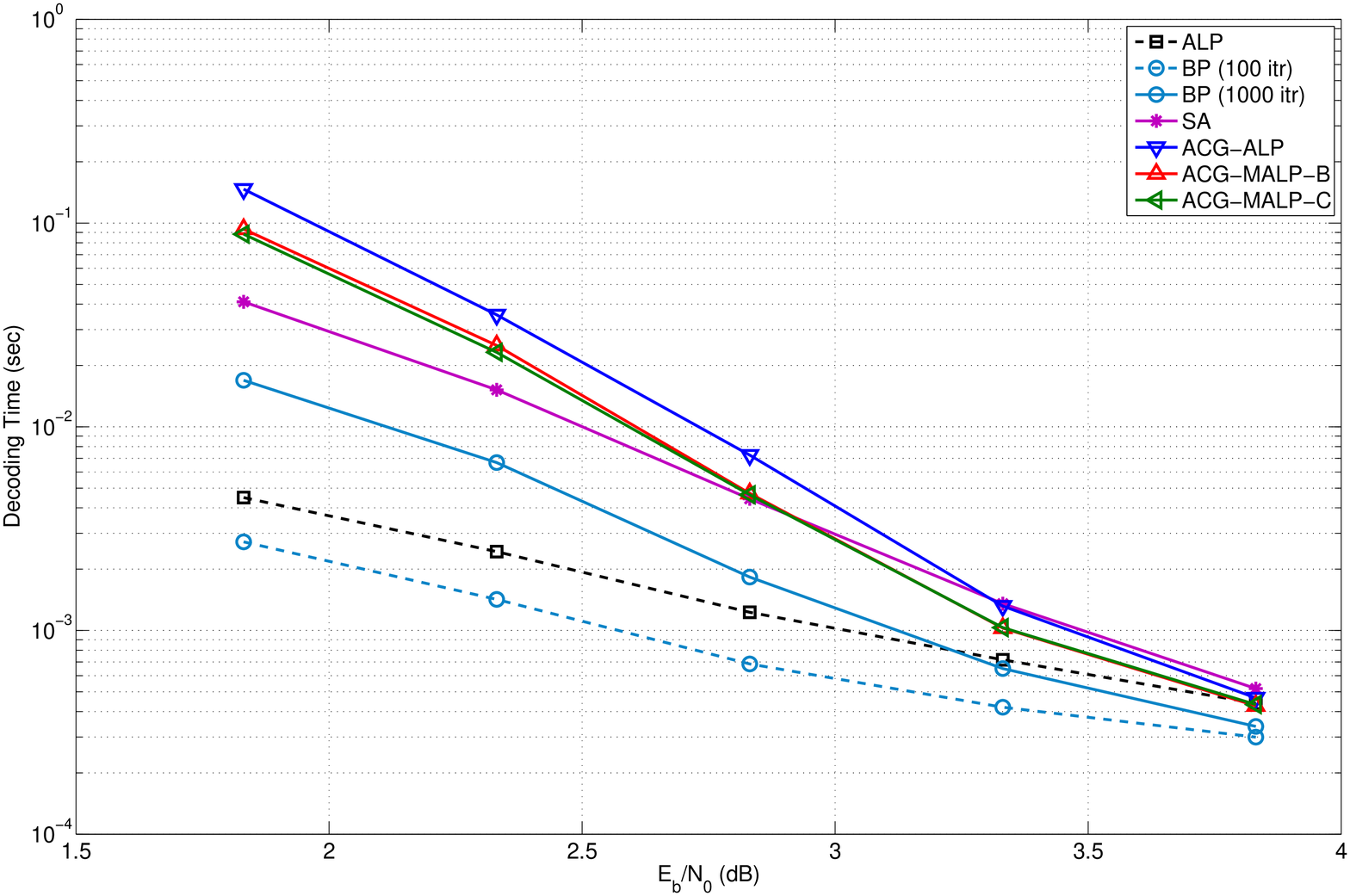}
\centering
\caption{Average simulation time for decoding one codeword of (155,64) Tanner LDPC code.}\label{fig_simt}
\end{figure}

From Fig.~\ref{fig_iter} and Fig.~\ref{subfig_cst}, we can see that, as expected, the ACG-ALP decoder takes fewer iterations to decode a codeword on average than the ACG-MALP-B/C decoders, but the ACG-MALP-B/C decoders have fewer constraints in each iteration, including the final iteration. We have observed that the ACG-MALP-B/C decoders require a larger number of iterations to decode than the ACG-ALP decoder, and fewer cuts are added into the constituent LP problems in each iteration on average, as reflected in Fig.~\ref{subfig_CpI}. This is because there are some iterations in which the added constraints had been previously removed.
Among all three proposed ACG-based decoding algorithms, we can see that the ACG-ALP decoder has the largest number of constraints in the final iteration and needs the least overall number of iterations to decode, while  ACG-MALP-B decoding has the smallest number of constraints but requires the largest number of iterations. The ACG-MALP-C decoder offers a tradeoff between those two: it has fewer constraints than the ACG-ALP decoder and requires fewer iterations than the ACG-MALP-B decoder. If we use the accumulated number of constraints in all iterations to decode one codeword as a criterion to judge the efficiency of these algorithms during simulation, then ACG-MALP-C decoding is more efficient than the other two algorithms in the low and moderate SNR regions, as shown in Table~\ref{tabCST}. Note that the ACG-MALP-B decoder is most efficient at high SNR where the decoding of most codewords succeeds in a few iterations and the chance of a previously removed inactive constraint being added back in later iterations is quite small. Hence, ACG-MALP-B decoding is preferred in the high-SNR region.

Fig.~\ref{fig_simt} presents an alternative way of comparing the complexity of the decoding algorithms. It shows the average decoding time when we implement the algorithms using C++ code on a desktop PC, with GLPK as the LP solver. The BP decoder is implemented in software with messages represented as double-precision floating-point numbers, and the exact computation of sum-product algorithm is used, without any simplification or approximation. The BP decoder iterations stop as soon as a codeword is found, or when the maximum allowable number of iterations -- here set to 100 and 1000 -- have been attempted without convergence. The simulation time is averaged over the number of transmitted codewords required for the decoder to fail on 200 codewords.

We observe that the ACG-MALP-B and ACG-MALP-C decoders are both uniformly faster than ACG-ALP over the range of SNR values considered, and, as expected from Table~\ref{tabCST}, ACG-MALP-C decoding is slightly more efficient than ACG-MALP-B decoding in terms of actual running time.
Of course, the decoding time depends both on the number of LP problems solved and the size of these LP problems, and the preferred trade-off depends heavily upon the implementation, particularly the LP solver that is used. Obviously, the improvement in error-rate performance provided by all three ACG-based decoding algorithms over the ALP decoding comes at the cost of increased decoding complexity. As SNR increases, however, the average decoding complexity per codeword of the proposed algorithms approaches that of the ALP decoder. This is because, at higher SNR, the decoders can often successfully decode the received frames without generating RPC cuts.

Fig.~\ref{fig_iter} shows that the ACG-ALP decoder requires, on average, more iterations than the SA decoder.
Our observations suggest that this is a result of the fact that the ACG-ALP decoder can continue to generate new RPC cuts after the number of iterations at which the SA decoder can no longer do so and, hence, stops decoding. The  simulation data showed that the additional iterations of the ACG-ALP decoder often resulted in a valid codeword, thus contributing to its superiority in perfomance relative to the SA decoder.

From Fig.~\ref{subfig_CpI}, it can be seen that the ACG-ALP-based decoding algorithms generate, on average, fewer cuts per iteration than the SA decoder. Moreover, as reflected in Fig.~\ref{subfig_Hcut} and \ref{subfig_Rcut}, the ACG-ALP decoders find more cuts from the original parity-check matrix and generate fewer RPC cuts per codeword. These observations suggest that the CSA is very efficient in finding cuts from a given parity check, while the SA decoder tends to generate RPCs even when there are still some cuts other than the Gomory cuts that can be found from the original parity-check matrix. This accounts for the fact, reflected in Fig.~\ref{fig_simt}, that the SA becomes less efficient as SNR increases, when the original parity-check matrix usually can provide enough cuts to decode a codeword. The effectiveness of our cut-search algorithm permits the ACG-ALP-based decoders to successfully decode most codewords in the high-SNR region without generating RPCs, resulting in better overall decoder efficiency.

Due to limitations on our computing capability, we have not yet tested our proposed algorithms on LDPC codes of length greater than 1000. We note that, in contrast to \cite{Taghavi_ALP} and \cite{SparseLP}, we cannot give an upper bound on the maximum number of iterations required by the ACG-ALP-based decoding algorithms because RPCs and their corresponding parity inequalities are generated adaptively as a function of intermediate pseudocodewords arising during the decoding process. Consequently, even though the decoding of short-to-moderate length LDPC codes was found empirically to converge after an acceptable number of interations, some sort of constraint on the maximum number of iterations allowed may have to be imposed when decoding longer codes. Finally, we point out that the complexity of the algorithm for generating cut-inducing RPCs lies mainly in the Gaussian elimination step, but as applied to binary matrices, this requires only logical operations which can be executed quite efficiently.

\section{Conclusion}
\label{sec:concl}
In this paper, we derived a new necessary condition and a new sufficient condition for a parity-check constraint in a linear block code parity-check matrix to provide a violated parity inequality, or cut, at a pseudocodeword produced by LP decoding. Using these results, we presented an efficient algorithm to search for such cuts and proposed an effective approach to generating cut-inducing redundant parity checks (RPCs). The key innovation in the cut-generating approach is a particular transformation of the parity-check matrix used in the definition of the LP decoding problem. By properly re-ordering the columns of the original parity-check matrix and transforming the resulting matrix into a ``partial'' reduced row echelon form, we could efficiently identify RPC cuts that were found empirically to significantly improve the LP decoder performance. We combined the new cut-generation technique with three variations of adaptive LP decoding, providing a tradeoff between the number of iterations required and the number of constraints in the constituent LP problems.  Frame-error-rate (FER) simulation results for several LDPC codes of length up to 999 show that the proposed adaptive cut-generation, adaptive LP (ACG-ALP) decoding algorithms outperform other enhanced LP decoders, such as the separation algorithm (SA) decoder, and significantly narrow the gap to ML decoding performance for LDPC codes with short-to-moderate block lengths.

\ifCLASSOPTIONcaptionsoff
  \newpage
\fi

\end{document}